\documentclass[preprint,11pt]{revtex4}

\usepackage[dvips]{graphicx}
\usepackage{amssymb,amsfonts,amsmath}

\newcommand{\be}{\begin{equation}}
\newcommand{\ee}{\end{equation}}

\newcommand{\bse}{\begin{subequations}}
\newcommand{\ese}{\end{subequations}}
\newcommand{\bea}{\begin{eqnarray}}
\newcommand{\eea}{\end{eqnarray}}

\newcommand{\bnum}{\begin{enumerate}}
\newcommand{\enum}{\end{enumerate}}

\newcommand{\bit}{\begin{itemize}}
\newcommand{\eit}{\end{itemize}}

\newcommand{\bc}{\begin{cases}}
\newcommand{\ec}{\end{cases}}

\newcommand{\bpm}{\begin{pmatrix}}
\newcommand{\epm}{\end{pmatrix}}

\newcommand{\bvm}{\begin{vmatrix}}
\newcommand{\evm}{\end{vmatrix}}

\newcommand{\bs}{\boldsymbol}

\newcommand{\mrm}{\mathrm}

\newcommand{\ga}{\alpha}

\newcommand{\gd}{\delta}
\newcommand{\eps}{\epsilon}

\newcommand{\gl}{\lambda}
\newcommand{\gk}{\kappa}

\newcommand{\gs}{\sigma}

\newcommand{\Go}{\Omega}
\newcommand{\Gc}{\Gamma}

\newcommand{\Gd}{\Delta}

\newcommand{\Gl}{\Lambda}

\newcommand{\gvf}{\varphi}

\newcommand{\diff}{\mrm{d}}

\newcommand{\lan}{\langle}
\newcommand{\ran}{\rangle}

\newcommand{\csp}{\;,\qquad}

\newcommand{\f}{\frac}

\begin{document}

\title{L\'evy Fluctuations and Tracer Diffusion in Dilute Suspensions of Algae and Bacteria}

\author{Irwin M. Zaid}
\affiliation{Rudolf Peierls Centre for Theoretical Physics, University of Oxford, 1 Keble Road, Oxford  OX1 3NP, United Kingdom}

\author{J\"orn Dunkel}
\affiliation{Rudolf Peierls Centre for Theoretical Physics, University of Oxford, 1 Keble Road, Oxford  OX1 3NP, United Kingdom}

\author{Julia M. Yeomans}
\affiliation{Rudolf Peierls Centre for Theoretical Physics, University of Oxford, 1 Keble Road, Oxford  OX1 3NP, United Kingdom}

\begin{abstract}
Swimming microorganisms rely on effective mixing strategies to achieve efficient nutrient influx.
Recent experiments, probing the mixing capability of unicellular biflagellates, revealed that passive
tracer particles exhibit anomalous non-Gaussian diffusion when immersed in a dilute suspension of self-motile
\emph{Chlamydomonas reinhardtii} algae. Qualitatively, this observation can be explained by the fact that the algae
induce a fluid flow that may occasionally accelerate the colloidal tracers to relatively large velocities. A satisfactory
quantitative theory of enhanced mixing in dilute active suspensions, however, is lacking at present. In particular, it
is unclear how non-Gaussian signatures in the tracers' position distribution are linked to the self-propulsion mechanism of a
microorganism. Here, we develop a systematic theoretical description of anomalous tracer diffusion in active suspensions, based
on a simplified tracer-swimmer interaction model that captures the typical distance scaling of a microswimmer's flow field. We show that
the experimentally observed non-Gaussian tails are generic 
and arise due to a combination of truncated L\'evy statistics for the velocity field and algebraically decaying time correlations in the fluid. Our analytical considerations are illustrated
through extensive simulations, implemented on graphics processing units to achieve the large sample sizes required for analyzing the tails
of the tracer distributions. 
\end{abstract}

\maketitle
%






Brownian motion presents one of the most beautiful manifestations of the
central limit theorem in Nature \cite{2005HaMa_R}. Reported as early as 1784 by the
Dutch scientist Jan van Ingenhousz \cite{1784IH}, the seemingly unspectacular random motion
of mesoscopic particles in a liquid environment made an unforeseeable impact when Perrin's
seminal experiments of 1909 \cite{1909Pe} yielded convincing evidence for the atomistic structure
of liquids. This major progress in our understanding of non-living matter -- which happened long before direct
observations of atoms and molecules came within experimental reach -- would not have been possible
without the works of Sutherland \cite{1905Su} and Einstein \cite{1905Einstein_BM}, who were able to
link the microscopic properties of liquids to a macroscopic observable, namely the mean square displacement
of a colloidal test particle.
\par
Caused by many quasi-independent random collisions with surrounding molecules, Brownian motion in a passive
liquid is quintessentially Gaussian, as predicted by the central limit theorem. Remarkably, however, recent
experiments by Leptos {\em et al.} \cite{2009LeEtAl_Gold} revealed notable non-Gaussian features in the
probability distribution of a tracer particle, when a small concentration of microscale swimmers, in their case
unicellular biflagellate \emph{Chlamydomonas reinhardtii} algae, was added to the fluid. Understanding this apparent
violation of the central limit theorem presents a challenging unsolved problem, whose solution promises new insights
into the mixing strategies of microorganisms \cite{1977Pu}. Here, we shall combine extensive analytic and large-scale numerical
calculations to elucidate the intimate connection between the flow field of an individual microorganism and the anomalous
(non-Gaussian) diffusion of tracer particles in a dilute swimmer suspension. 
\par
Modern high-speed microscopy techniques resolve the stochastic dynamics of micron-sized tracer particles
to an ever increasing accuracy \cite{2006Lubensky,2010Li}. This opens the exciting possibility of using high-precision
tracking experiments to probe the statistics of the flow fields created by active swimmers, and hence their connection
to physical properties and evolutionary strategies of microorganisms that live in liquid environments \cite{2009PoEtAl_Gold,2010Drescher_PNAS,2009LeEtAl_Gold}.
Furthermore, a novel class of micromechanical devices \cite{2009So, 2010Le} use nonequilibrium fluctuations
generated by bacteria as a fundamental ingredient of their operation. To explain and exploit the nonequilibrium
conditions in active suspensions, it is important to fully understand the relation between the experimentally observed
features of tracer displacements and the characteristics (such as self-propulsion mechanisms) of the algae or bacteria.
\par
The observations of Leptos {\em et al.} \cite{2009LeEtAl_Gold} demonstrate that the time-dependent probability distributions 
of tracer displacements in dilute algae suspensions exhibit tails that decay much more slowly than would be expected if the
tracers obeyed purely Gaussian statistics. At high swimmer concentrations, enhanced transport might be expected
\cite{2000WuLi_PRL,2007LuYo_PRL,2005HeOr_PRL, 2008UnHe_PRL}, as collective behavior emerges from swimmer interactions, which can
lead to the formation of large-scale vortices and jets. In dilute suspensions, however, where swimmer-swimmer interactions can be neglected,
a satisfactory quantitative understanding of the underlying velocity statistics is still lacking. Below, we are going to show that the velocity
distribution produced by the swimmers takes a tempered (or regularized) L\'evy form \cite{2000MeKl, 2004MeKl}, and that the long-time behavior of the tracers' positional probability
distribution function can be understood in terms of correlated truncated L\'evy flights \cite{1994Stanley}.
\par
If a microswimmer is self-propelled, with no external forces acting, its flow field scales with distance as $r^{-n}$ for an
exponent $n\ge 2$ \cite{2010DrEtAl_PRL}. We will demonstrate that it is this form of the power-law decay that is responsible for the anomalous diffusion of tracer particles~\cite{2009LeEtAl_Gold}. Remarkably, qualitatively different behavior can be expected in suspensions of sedimenting swimmers: If gravity plays an important role for the swimmer dynamics, the  far field flow decays as $r^{-1}$ and tracers will diffuse normally. Finally, our results suggest that, on sufficiently long times scales, anomalous tracer diffusion in dilute active suspensions can be viewed
as a natural example of a stochastic process described by a fractional diffusion equation.

\section{Model}

Given an advecting flow $\bs{u}_N(t,\bs{r})$, generated by a dilute suspension of $N$
self-swimming microorganisms, we model the dynamics $\bs x(t)$ of a passive, colloidal
tracer particle (radius $a$) by an overdamped Langevin equation of the form
\begin{equation}\label{e:LE}
\frac{d}{dt} \bs x(t) = \bs{u}_N(t, \bs{x}(t)) + \sqrt{2D_0}\boldsymbol{\xi}(t).
\end{equation}
The random function $\bs{\xi}(t)=(\xi_i(t))_{i=1,2,3}$ represents uncorrelated Gaussian white noise
with $\langle \xi_{i}(t) \rangle = 0$ and $\langle \xi_{i}(t) \xi_{j}(t')\rangle = \delta_{i j} \delta(t - t')$,
describing stochastic collisions between the tracer and surrounding liquid molecules. The thermal diffusion coefficient
$D_0$ in a fluid of viscosity $\eta$ is determined by the Stokes-Einstein relation $D_0 = k_{\mathrm{B}} T / (6 \pi \eta a)$.
\par
If the Reynolds number is very small, the net flow due to $\gs =1,\ldots, N$ active swimmers, located at positions $\bs X^\gs(t)$
and moving at velocity $\bs V^\gs (t)$, is, in good approximation, the sum of their individual flow fields $\bs u$,
\begin{equation}\label{e:flow}
\bs u_N(t, \bs x)=\sum_{\gs=1}^N\bs u(\bs x|\Gc^\gs(t))
\csp
\Gc^\gs(t):=(\bs X^\gs(t),\bs V^\gs (t)).
\end{equation}
Since we are interested in physical conditions similar to those in the experiments of Leptos {\em et al.}~\cite{2009LeEtAl_Gold}, our analysis will
focus on a dilute suspension of active particles, corresponding to the limit of a small volume filling fraction $\gvf\ll 1$. In this case, binary
encounters between swimmers are negligible perturbations. Moreover, we can ignore random reorientation of swimmers caused by rotational diffusion
(due to thermal fluctuations) and search behavior (like chemotaxis or phototaxis), since these effects take place on the order of several seconds and
thus are not relevant to the tracer dynamics. Indeed, in dilute homogeneous solutions, it is irrelevant for the tracer statistics (even on longer time scales) 
whether a tracer experiences two successive scatterings from the same tumbling swimmer or from two different non-tumbling swimmers. It is therefore sufficient
to assume that each swimmer moves ballistically, 
\be
\bs X^\gs(t)=\bs X_0^\gs+t {\bs V}^\gs_0.
\ee
For dilute suspensions, the initial swimmer coordinates $\Gc^\gs(0)=(\bs X^\gs_0,\bs V^\gs_0)$ are independent and identically distributed random
variables with one-particle PDF $\Phi_1(\Gc^\gs_0)$. More specifically, we assume that the distribution of the initial positions $\bs X^\gs_0$ is
spatially uniform and that the swimmers have approximately the same speed $V$, that is $\Phi_1(\Gc^\gs_0)\propto \gd(V-|\bs V_0^\gs|)$. 
\par
To complete the definition of the model, we need to specify the flow field $\bs u(\bs x|\Gc^\gs(t))$ generated by a single swimmer. There are various
strategies for achieving directed propulsion at the microscale \cite{2009LaPo}. Small organisms, like algae and bacteria, can swim by moving slender
filaments in a manner not the same under time reversal. Self-motile colloids, a class of miniature artificial swimmers, are powered with interfacial
forces induced from the environment \cite{1989An}. Although both of these are active particles, microscopic details of their geometry and self-propulsion can lead
to different velocity fields. This, in turn, affects how a tracer migrates in their flow.  In the Stokes regime, if external force are absent,  self-propelled particles or microorganisms generate velocity fields decaying as $r^{-2}$ or
faster~ \cite{2010DrEtAl_PRL,1989An}. Since we are interested in the general features of mixing
by active suspensions, and there is no universal description of the flow around an active object, it is helpful to consider simplified velocity field models
that capture generic features of real microflows.
\begin{figure}
\centering
\includegraphics{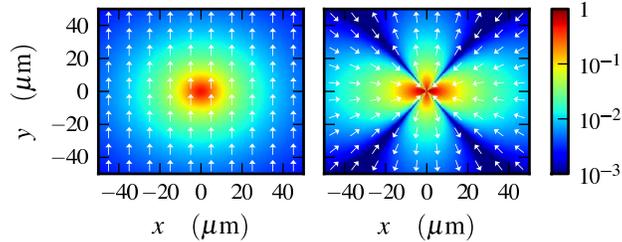}
\caption{Flow fields of the model swimmers considered in our analytical and numerical
calculations. In both plots, the swimmer is oriented upwards and the flow is normalized
by the swimming speed. The arrows indicate velocity direction and the colors represent magnitude.
(Left) Co-oriented model with $r^{-2}$ decay and (Right) stroke-averaged flow generated by a dipolar ``pusher''.
Parameters are those of Fig.~\ref{fig_vel_PDF}.
\label{fig_model}}
\end{figure}
\par
We shall focus on two simplified models  (see Fig.~\ref{fig_model})  
that can be interpreted as contributions in a general multipole expansion of a flow field. Specifically,
we will compare a co-oriented model~\cite{2010DuPuZaYe} with trivial angular dependence 
\bse\label{e:models}
\be\label{e:toy_model}
\bs u(\bs x| \Gc^\gs)
= (\gk V)\;\left(\f{\eps^n}{|\bs R^\gs|^n+\gl^n}\right)\hat{\bs \Go}^\gs
\csp
n\ge 1 
\ee
to a more realistic dipolar (or stresslet) flow field \cite{1990PeKe} 
\be\label{e:dipol_model}
\bs u(\bs x| \Gc^\gs)
= 
(\gk V) \left(\f{  \eps^2}{|\bs R^\gs|^2+\gl^2}\right)
\left[
3\left(\hat{\bs \Go}^\gs \hat{\bs R}^\gs\right)^2-1
\right]
\hat{\bs R}^\gs.
\ee
\ese
In Eqs.~\eqref{e:models}, the vector $\bs R^\gs(t):=\bs X^\gs(t)-\bs x$ connects the swimmer and
tracer position at time $t$, $\hat{\bs R}^\gs(t):=\bs R^\gs/|{\bs R}^\gs|$ is the associated unit
vector, and $\hat{\bs \Go}^\gs:=\hat{\bs V}^\gs/V$ defines the swimmer's orientation and swimming
direction. The parameter $\eps$ characterizes the swimmer length scale, $\kappa$ is a dimensionless
constant that relates the amplitude of the flow field to the swimmer speed, and $\gl$ regularizes the
singularity of the flow field at small distances. The co-oriented model~\eqref{e:toy_model}, due to its minimal angular dependence, is useful for pinpointing how the tracer statistics depend on the distance scaling
of the flow field. For $n=1$, the scaling is equivalent to that of an ``active'' colloid or forced swimmer, whereas for
$n\ge 2$ the scaling resembles that of various natural swimmers not subjected to an external force. In
particular, the case $n=2$ allows us to ascertain the effects of the angular dependence of the flow field structure
on tracer diffusion, by comparing against the more realistic dipolar model \eqref{e:dipol_model}. The latter is commonly
considered as a simple stroke-averaged description for natural microswimmers \cite{2009BaMa_PNAS, 2002SiRa_PRL}. As shown in Ref.~\cite{2010DuPuZaYe},
stroke-averaged models are able to capture the most important aspects of the tracer dynamics on time scales longer than the swimming
stroke of a microorganism.

\section{Results}
We are interested in computing experimentally accessible, statistical properties of the tracer particles, such as their velocity PDF,
correlation functions, and position PDF. These quantities are obtained by averaging suitably defined functions with respect to the $N$-swimmer
distribution $\Phi_N=\prod_\gs \Phi_1(\Gc^\gs_0)$. A detailed description of the averaging procedure and a number of exact analytical
results are given in the Supplementary Material, below we shall restrict ourselves to discussing the main results and their implications. 
\par
We begin by considering the equal-time velocity PDF and velocity autocorrelation function at a fixed point in the fluid. Since we are
primarily interested in the swimmer contributions, we will focus on the deterministic limit $D_0=0$ first. The additive effect of thermal
Brownian motion will be taken into account later, when we discuss the position statistics of the tracer. Considering a suspension of $N$ swimmers,
confined by a spherical volume of radius $\Gl$, the equal-time velocity PDF $\phi_{N,\Gl}(\bs v)$ and velocity autocorrelation function $C_{N,\Gl}(t)$
of the flow field near the center of the container are formally defined by 
\bse
\begin{eqnarray}
\label{e:def_vPDF}
\phi_{N,\Gl}(\bs v) :=\lan \gd(\bs v-\bs u_N(0,\bs 0))\ran,  
\\
C_{N,\Gl}(t) := \lan \bs u_N(t,\bs 0) \bs  u_N(0,\bs 0)\ran,
\end{eqnarray}
\ese
where the average $\lan\; \cdot\;\ran$ is taken with respect to the spatially uniform initial distribution of the swimmers.
For the models~\eqref{e:models}, it is possible to determine $\phi_{N,\Gl}$ and $C_{N,\Gl}$ analytically.

\subsection{Velocity PDF: Slow convergence of the central limit theorem for $\boldsymbol{n \geq 2}$}

To elucidate the origin of the unusual velocity statistics in an active suspension, let us consider the tracer
velocity PDF when there is a single swimmer present, $\phi_{1,\Lambda}(\bs v)$. The tail of this function reflects
large tracer velocities generated by close encounters with the swimmer. It is instructive to start with the (unphysical)
limit $\lambda=0$, where the interaction diverges at short distances and there is no cutoff for large velocities. In this case,
one readily finds from \eqref{e:def_vPDF} that asymptotically $\phi_{1,\Gl}(0,\bs v)\propto |\bs v|^{-(3+3/n)}$. This means that
the variance of the probability distribution is finite for $n=1$, but infinite for $n\ge 2$. According to Eq.~\eqref{e:flow}, the
flow field due to $N$ swimmers is the sum of $N$ independent and identically distributed random variables. Hence, the central limit
theorem predicts that, for $\gl=0$ and $n=1$, the velocity distribution $\phi_{N,\Gl}$ converges to a Gaussian in the large $N$ limit,
whereas for $\gl=0$ and $n\ge 2$ one expects non-Gaussian behavior due to the infinite variance of $\phi_{1,\Gl}$.
\par
For a real swimmer, the flow field is strongly increasing in the vicinity of the swimmer \cite{2010DrEtAl_PRL, 2010GuGo}, but remains finite
due to lubrication effects and nonzero swimmer size. This corresponds in our model to a positive value of $\lambda$. For $\gl>0$, the variance of
the one-swimmer PDF $\phi_{1,\Gl}$ remains finite and, formally, the conditions for the central limit theorem are satisfied for all $n\ge 1$. However,
for $n\ge 2$ the variance of $\phi_{1,\Gl}$ remains very large and the convergence to a Gaussian limiting distribution is very slow. Our subsequent analysis
demonstrates that the velocity PDF is more accurately described by a tempered L\'evy-type distribution.
\par
\begin{figure*}
\centering
\includegraphics{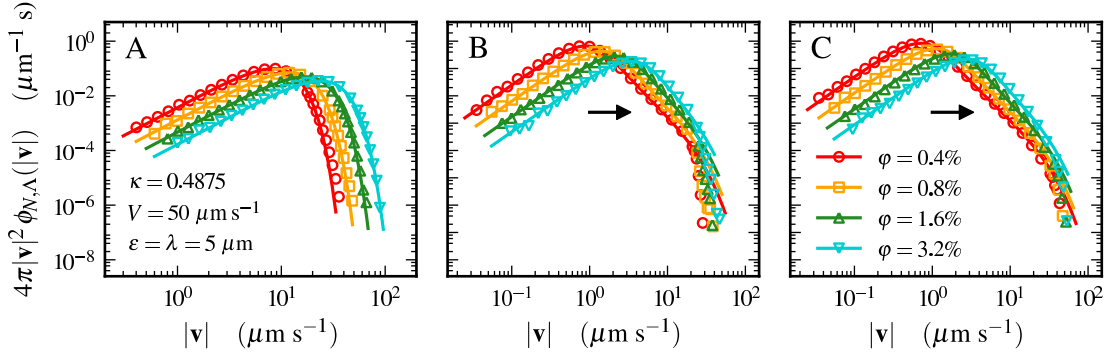}
\caption{Velocity PDF of a tracer particle in the flow generated by different concentrations of swimmers.
The solid curves are based on approximation \eqref{e:ansatz_irwin}, using the exact second and fourth moments
of the velocity PDFs, as shown in the Supplementary Material. (A) For the co-oriented model \eqref{e:toy_model}
with long-range hydrodynamics $n=1$, the velocity PDFs from simulations (symbols) converge rapidly to the Gaussian
distribution predicted by the central limit theorem (solid curves), even at low volume fractions $\gvf\simeq 0.4\%$. (B) In contrast,
for the co-oriented model with $n=2$, the central limit theorem convergence is very slow and the velocity distribution exhibits strongly
non-Gaussian features at volume fractions similar to those realized by recent experiments \cite{2009LeEtAl_Gold}. (C) The velocity PDF for the dipolar swimmer
model looks very similar to that of our co-oriented model (B), which means the angular dependence does not play an important role for
the velocity distribution. Simulation parameters are given in A, the sample size is $4 \times 10^{6}$ throughout.
\label{fig_vel_PDF}
}
\end{figure*}

These statements are illustrated by Fig.~\ref{fig_vel_PDF}, which shows velocity PDFs obtained numerically (symbols)
and from analytical approximations (solid curves) for the co-oriented model with $n=1$ (A) and $n=2$ (B), and the dipolar model (C) at different swimmer
volume fractions $\gvf:=N(\eps/\Gl)^3$. As evident from Fig.~\ref{fig_vel_PDF}~A, for $n=1$ the velocity PDF indeed converges rapidly to
the Gaussian shape, in accordance with the central limit theorem. By contrast, for velocity fields decaying as $r^{-n}$ with $n \ge 2$, the
convergence is surprisingly slow and one observes strongly non-Gaussian features at small filling fractions~$\gvf$. The arrows highlight
this regime, which shows a power-law dependence of the velocity distribution on the magnitude of the velocity, a signature of a L\'evy distribution.
\par
The remarkably slow convergence to the Gaussian central limit theorem prediction can be understood quantitatively by considering the characteristic function 
\be
\chi_\phi(\bs q)
:=
\int\diff^3 v \;
\exp(-i\bs q \bs v)\;\phi(\bs v)
\ee
of the velocity PDF~\eqref{e:def_vPDF}. A detailed analytical calculation (see Supplementary Material) shows that the exact result
for $\chi_\phi(\bs q)$ can be approximated by 
\be\label{e:ansatz_irwin}
\tilde{\chi}_\phi(|\bs q|)=\exp\{-[(c\; |\bs q|^2+\mu^2 )^{\ga/2}-\mu^\ga]\}.
\ee
This expression, which reduces to a Gaussian for $\ga=2$, is of the tempered L\'evy form, 
and gives rise to the following tracer velocity moments 
\bse\label{e:ansatz_irwin_moments}
\bea
\lan  |\bs v|^2\ran
&=&
3 \ga c\;   \mu ^{\ga-2},
\\
\lan  |\bs v|^4\ran
&=&
15 \ga c^2  \mu ^{\ga-4} \left[2+\ga  \left(\mu^{\ga }-1\right)\right].
\eea
\ese
By studying asymptotic behavior in the small cutoff limit $\gl\to 0$ one finds that, for
velocity fields $\bs u$ decaying as $r^{-n}$,
\be
n=1 \quad\Rightarrow\quad \ga=2
\csp\quad
n\ge 2 \quad\Rightarrow\quad  \ga=3/n.
\ee
This result confirms that for colloidal-type interactions with $n=1$ the velocity PDF is Gaussian,
whereas for $n\ge 2$ deviations from Gaussianity occur in agreement with our numerical results of
Fig.~\ref{fig_vel_PDF}. In the limit $\mu=0$, Eq.~\eqref{e:ansatz_irwin} describes the family of L\'evy stable
distributions. These distributions arise from a generalized central limit theorem \cite{1954GnKo} relevant to
random variables having an infinite variance. Specifically for $n=2$, one recovers the Holtsmark distribution that describes
the statistics of the gravitational force acting on a star in a cluster \cite{1943Ch} and of the velocity field created
by point vortices in turbulent flows \cite{1984Ta}.
\par
However, for realistic non-singular flow fields, corresponding to finite values of $\gl$, 
we generally have $\mu>0$. Specifically, by matching the exact velocity moments to those
in Eq.~\eqref{e:ansatz_irwin_moments}, one finds that for the co-oriented model~\eqref{e:toy_model}
with $n=1$ 
\be\label{e:sigma_1}
c=\f{1}{2} N \left(\f{\eps}{\Gl}\right)^2\ (\gk V)^2\;\left[1+\ell+2\ell
\log(\ell)\right]
\ee
at leading order in $\ell:=\gl/\Gl$, whereas for $n=2$
\bse\label{e:mu_2}
\bea
c
&=&
\f{1}{3} \left(\frac{5\pi^4}{12}\right)^{1/3}
\left(\kappa V\right)^2
\gvf^{4/3},
\\\
\mu
&=&
\left(\frac{10 \pi }{3}\right)^{2/3} \left(\f{\gl}{\eps}\right)^2 \gvf^{2/3}.
\label{e:mu_2b}
\eea
\ese
For the dipolar model~\eqref{e:dipol_model}, one obtains the same scaling of $(c,\mu)$
with $(\gvf,\gl)$ as in Eq.~\eqref{e:mu_2} but with a slightly different numerical
prefactor, yielding in the small cutoff limit $\gl\to 0$
\bse\label{e:dipol_moments_approx}
\bea
\label{e:dipol_moments_approx_a}
\lan |\bs v|^2\ran
&=&
\f{3\pi}{5}(\gk V)^2 \;\gvf
\left(\f{\eps}{\gl}\right),
\\
\lan |\bs v|^4\ran
&=&
\f{9\pi}{70}(\gk V)^4 \;\gvf
\left(\f{\eps}{\gl}\right)^5.
\eea
\ese
Note that Eq.~\eqref{e:sigma_1} suggests for colloidal-type flow fields with $\bs
u\propto r^{-1} $ the appropriate thermodynamic limit is given by $N,\Gl\to\infty$ such that
$N/\Gl^2=\mathit{const.}$, whereas we must fix $\gvf=\mathit{const.}$ if $\bs u\propto r^{-n}, n\ge 2$.
Furthermore, Eqs.~\eqref{e:mu_2b} and~\eqref{e:dipol_moments_approx_a} imply that $\mu\to
0$ and $\lan |\bs v|^2\ran\to\infty$ for a vanishing regularization parameter $\gl\to 0$.
This illustrates that L\'evy-type behavior becomes more prominent the more ``singular'' the velocity field in the vicinity of the swimmer. 
\par
The solid curves in Fig.~\ref{fig_vel_PDF} are based on approximation \eqref{e:ansatz_irwin}, using the
exact second and fourth moments of the velocity PDFs, as given in the Supplementary Material. For $\bs u\propto r^{-2} $,
the Gaussian prediction of the central limit theorem becomes accurate only at large volume fractions ($\gvf \gtrsim
25\%$). In the dilute regime $\gvf\ll1 $, the bulk of the probability comes from a L\'evy stable distribution before
it crosses over to quasi-Gaussian decay, reflected by the (truncated) power-law tails in Fig.~\ref{fig_vel_PDF} B and C. We may thus
conclude that the fluid velocity in a dilute swimmer suspension is a biophysical realization of truncated L\'evy-type random variables~\cite{1994Stanley}.

\subsection{Flow field autocorrelation}
\begin{figure}
\centering
\includegraphics{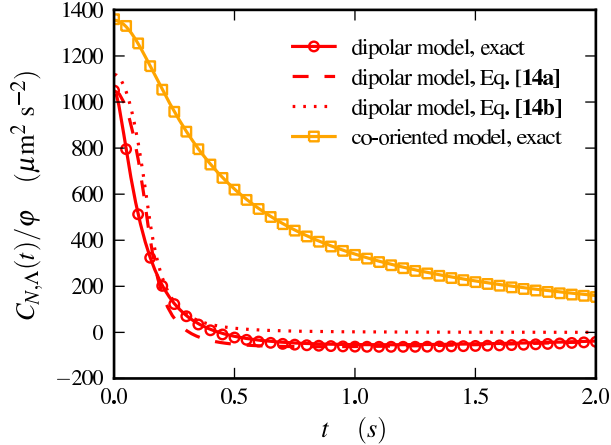}
\caption{Due to the different flow topologies, the velocity autocorrelation function for dipolar
swimmers \eqref{e:dipol_model} decays faster than that for the co-oriented model \eqref{e:toy_model} with
$n=2$. Solid curves indicate the exact analytic solution and symbols correspond to simulation data. Dotted and
dashed curves illustrate respectively the long time approximation and its behavior in the thermodynamic limit.
Parameters are those of Fig.~\ref{fig_vel_PDF}.
\label{fig_correlations}
}
\end{figure}

The similarity of  Figs.~\ref{fig_vel_PDF} B and C suggests that the angular flow field
structure is not important for the equal-time velocity distribution. By contrast, the
velocity autocorrelation function $C_{N,\Lambda}(t)$ depends sensitively on the angular details, as illustrated in
Fig.~\ref{fig_correlations}. For both our co-oriented model \eqref{e:toy_model} with $n=1,2$
and the dipolar model \eqref{e:dipol_model}, the function $C_{N,\Lambda}(t)$ can be determined
analytically (see Supplementary Material). From the exact results, one finds that for $n = 2$
in the thermodynamic limit at long times $t\gg \tau_\eps:=\epsilon/V$ 
\be\label{e:alpha=2_C_longtime}
C_{N,\Lambda}(t)\simeq\frac{3 \pi ^2}{4}  \varphi \,(\kappa V)^2\,  \,\left(\f{\tau_\eps}{t}\right).
\ee
For comparison, the velocity autocorrelation function
for dipolar swimmers can be approximated by 
\bse
\be\label{e:dip_C_approx_3}
C_{N,\Lambda}(t)\simeq
\gvf\;(\gk V)^2 \left(\f{\eps}{\Gl}\right)
\f{12}{5}
\begin{cases}
-1+\frac{1}{\ell_*}+\frac{3 s^2}{7}-\frac{3 s^2}{7 \ell_*^3}, &\ell_*\ge s,\\
-1-\frac{3 \ell_*^4}{7 s^5}+\frac{\ell_*^2}{s^3}+\frac{3 s^2}{7}, &\ell_*<s,
\end{cases}
\ee
where $\ell_*:={4\ell}/{\pi }$ and $s:=t V/\Gl<1$.
The approximation~\eqref{e:dip_C_approx_3}, shown as the dotted line in
Fig.~\ref{fig_correlations},  becomes exact at long times. In the thermodynamic limit,
it reduces to
\be\label{e:dip_C_approx_TDL}
C_{N,\Lambda}(t)\simeq
\gvf\;(\gk V)^2 \left(\f{\eps}{\gl}\right)
\f{3\pi}{5} 
\begin{cases}
1- \frac{3 t^2}{7 \tau_\gl^2}, & t  \leq \tau_\gl, \\
 \frac{ \tau_\gl^3}{t^3} - \frac{3 \tau_\gl^5}{7t^5}, & t> \tau_\gl,
\end{cases}
\ee
\ese
where $\tau_\gl:= 4\gl/(\pi V)$. Note that Eq.~\eqref{e:dip_C_approx_TDL} predicts an
asymptotic $t^{-3}$ decay, which is considerably faster than the $t^{-1}$ decay for the co-oriented 
model, cf. Eq.~\eqref{e:alpha=2_C_longtime}. This is due to the different angular structure
of their respective flow fields. The excellent agreement between simulation data and the exact analytic
curves (solid) in Fig.~\ref{fig_correlations} also confirms the validity of our simulation scheme
(see Numerical Methods in Supplementary Material).

\subsection{Mean square displacement} 
\begin{figure}
\centering
\includegraphics{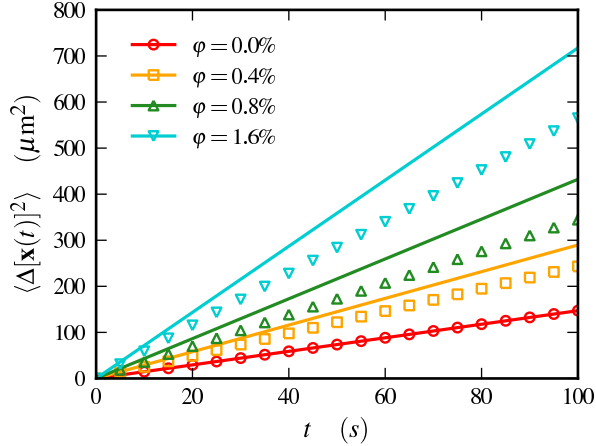}
\caption{Mean square displacement at different volume fractions. Solid lines are analytic
upper bounds, $  6 (D_{0} + \varphi \kappa^{2} V \epsilon) t\ge \lan\Gd[\bs x(t)]^2\ran$, and 
symbols display the ensemble average from simulations. Parameters are those of Fig.~\ref{fig_vel_PDF}.
\label{fig_msd}
}
\end{figure}

Having discussed the velocity statistics, we next analyze the tracer displacements. To that end,
we will focus on the practically more relevant dipolar swimmer case, and include the effects of
thermal Brownian motion (so $D_0>0$). A first quantifier, that can be directly measured in experiments,
is the mean square displacement $\lan \Gd[\bs x(t)]^2\ran\simeq \lan \Gd[\bs x(t)]^2\ran_N+6D_0t$.
Assuming spatial homogeneity and spatially decaying correlations, the velocity autocorrelation function
can be used to obtain an upper bound for the swimmer contribution
\bea
\lan \Gd[\bs x(t)]^2\ran_N
&=&\int_0^t\diff \tau' \int_0^t\diff\tau \;\lan \bs u_N(\tau',\bs x(\tau'))\bs u_N(\tau,\bs
x(\tau))\ran
\notag\\
&\le&\label{e:MSD_bound}
\int_0^t\diff \tau' \int_0^t\diff\tau \;\lan \bs u_N(\tau',\bs 0)\bs u_N(\tau,\bs 0)\ran.
\eea
Inserting the  approximate result~\eqref{e:dip_C_approx_TDL}, we find
\begin{eqnarray}\label{e:MSD}
\lan \Gd[\bs x(t)]^2\ran_N
\le
6\,\gvf\; \gk^2 V\eps\; t
\begin{cases}
\frac{2 t  }{5\tau_\gl}-\frac{t^3 }{35 \tau_\gl^3}, &  t<\tau_\gl,\\
1- \f{\tau_\gl}{t}+\frac{2  \tau_\gl^2}{5 t^2 }-\frac{\tau_\gl^4}{35 t^4},   & 
t\ge\tau_\gl.
\end{cases}
\end{eqnarray}
Equation~\eqref{e:MSD} implies that tracer diffusion due to the presence of dipolar
swimmers is ballistic at short times $t\ll \tau_\gl$ and normal at large times $t\gg
\tau_\gl$ (see Fig.~\ref{fig_msd}). Qualitatively, the predicted linear growth $\lan
\Gd[\bs x(t)]^2\ran\propto t$ agrees with the experimental results of Leptos et
al.~\cite{2009LeEtAl_Gold}. Generally, the asymptotic diffusion constant will be of the
form $D\simeq D_0+\nu \gvf\; \gk^2 V\eps$, where $\nu$ is a  numerical prefactor of order
unity that encodes spatial correlations neglected in Eq.~\eqref{e:MSD_bound}.

\subsection{Evolution of the position PDF}
The spatial motion of a passive tracer in a fluctuating flow $\bs{u}(t,\bs{r})$ is
described by the position PDF $P(t,\bs{r})=\lan\gd(\bs r-\bs x(t))\ran$. For Gaussian fields,
uniquely defined by the two-point correlation function $\langle u_{i}(t,\bs{r}) u_{j}(t',\bs{r}') \rangle$,
it is possible to characterize $P(t,\bs{r})$ analytically for some classes of trajectories $\bs x(t)$ \cite{1995HaJu}.
However, our analysis above indicates that the statistics in an active suspension are neither $\gd$-correlated nor Gaussian,
exhibiting features of L\'evy processes.
\par
Generally, the hierarchy of correlations in L\'evy random fields is poorly understood \cite{1994SaTa}. It is therefore unclear how
to adapt successful models of random advection by a Gaussian field \cite{1994ShSi} or, more generally, extend the understanding of
colored Gaussian noise \cite{1995HaJu} to colored L\'evy processes. These theoretical challenges make it very difficult, if not impossible,
to construct an effective diffusion model that bridges the dynamics of $P(t,\bs{r})$ on all of the time scales. Partial theoretical insight
can be gained, however, by considering the asymptotic short and long time behavior.
\par
\begin{figure}
\centering
\includegraphics{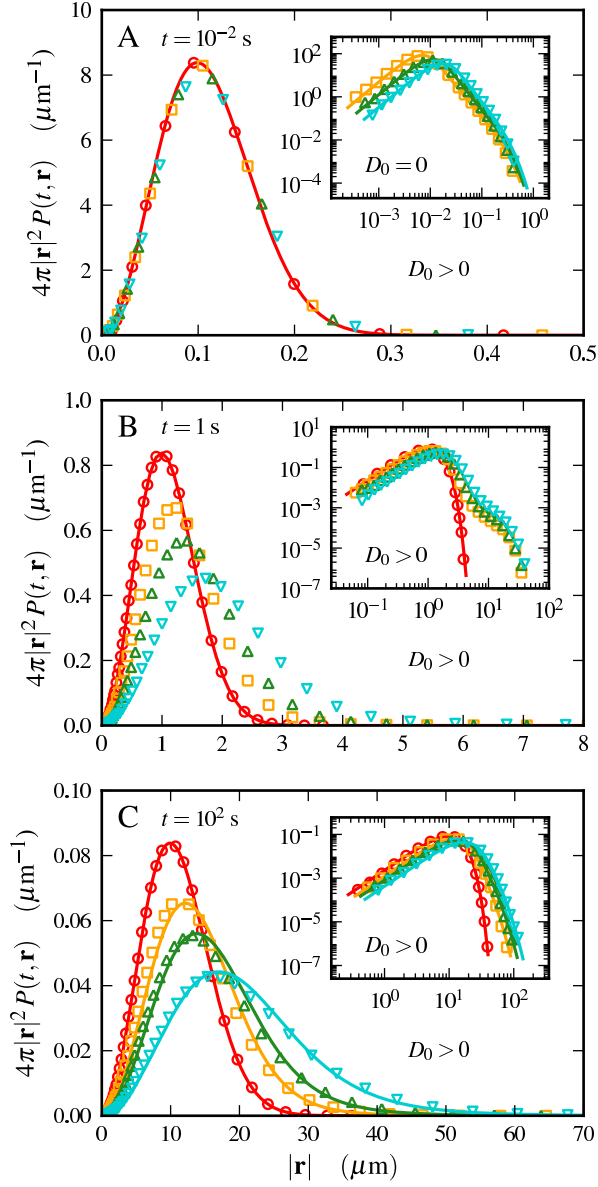}
\caption{Radial position PDF of a tracer in a dilute suspension of dipolar swimmers at various times. Solid curves
represent analytic forms of $P(t,\bs{r})$ and symbols illustrate histograms determined from simulations. Insets B and C
show the same quantities on a log-log scale. Volume fractions and symbols are those of Fig.~\ref{fig_msd}.
Parameters are those of Fig.~\ref{fig_vel_PDF} with $D_0 = 0.245 \ \mu \mathrm{m}^{2} \ \mathrm{s}^{-1}$. (A) Short time regime. At
these times, Brownian motion effectively dominates constant advection for our choice of parameters. In the limit of no thermal noise
(inset), the position PDF is the tempered L\'evy velocity PDF $\phi_{N,\Gl}(\bs v)$ after a rescaling with $t$. (B) Transient behavior. This
period corresponds to an intermediate decay of the velocity autocorrelation. (C) Asymptotic long time regime. Eventually, random advection
from many low Reynolds number swimmers becomes equivalent to a tempered L\'evy flight. The solution to the fractional diffusion
equation \eqref{e:frac} is matched against simulations by fitting its coefficients $D_{\alpha}$ and $K$, see also Fig.~\ref{fig_pos_pdf_1D}.
\label{fig_pos_pdf}}
\end{figure}

At short times $t \ll \tau_{\lambda}$, the position PDF combines ballistic transport from constant swimmer advection
and diffusive spreading from thermal Brownian effects. For experimentally-relevant parameters~\cite{2009LeEtAl_Gold}, 
normal diffusion is much stronger than the advection and, at these times, the dynamics of $P(t,\bs{r})$ are captured by
the normal diffusion equation. If Brownian motion is neglected, we have $P(t,\bs{r}) = t^{-3} \phi_{N,\Gl}(\bs r / t)$ with $\phi_{N,\Gl}(\bs v)$
as the tempered L\'evy velocity PDF. The resulting ``ballistic'' L\'evy distribution is in good agreement with simulation data for $D_0=0$ at
short times, see inset of Fig.~\ref{fig_pos_pdf}~A.
\par
\begin{figure}
\centering
\includegraphics{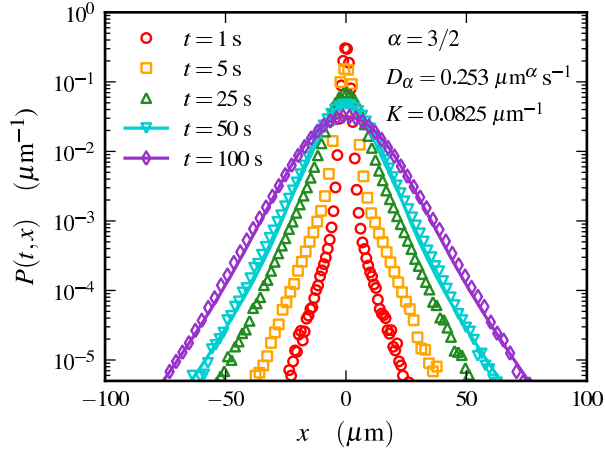}
\caption{Time evolution of the marginal position PDF at a volume fraction $\varphi = 1.6\%$. Solid curves
represent Eq.~\eqref{e:sol} with fitted coefficients (shown only at long times) and symbols illustrate histograms
determined from simulations. Parameters are those of Fig.~\ref{fig_vel_PDF} with $D_0 = 0.245 \ \mu \mathrm{m}^{2} \ \mathrm{s}^{-1}$.
At intermediate times $t \simeq 1 \ \mathrm{s}$, our data resemble the measurements from Ref.~\cite{2009LeEtAl_Gold}. 
\label{fig_pos_pdf_1D}}
\end{figure}

For long times $t \gg \tau_{\lambda}$, after the correlations of the velocity field have vanished (typically after several seconds),
we may interpret a tracer diffusing in an active suspension as a realization of an uncorrelated tempered L\'evy process. Effectively, this
corresponds to replacing $\bs{u}_N(t,\bs{r})$ from Eq.~\eqref{e:LE} with a $\gd$-correlated but non-Gaussian random function $\boldsymbol{\zeta}(t)$.
To characterize the statistical properties of the swimmer-induced noise $\boldsymbol{\zeta}(t)$, we need to specify its characteristic functional
$\mathcal{F}[t;\bs{k}(s)]$~\cite{2009TouchetteCohen}. Our earlier findings, that the asymptotic mean square displacement grows linearly in time and
that the velocity field amplitudes follow a tempered L\'evy stable distribution, suggest the functional form
\begin{eqnarray}
\label{e:noise}
\mathcal{F}[t;\bs{k}(s)] = 
\exp \left\{ D_{\alpha}  K^{\alpha} t - D_{\alpha} \int_{0}^{t} \diff s \, [K^{2} +
|\bs k(s)|^{2}]^{\alpha / 2} \right\}.
\end{eqnarray}
Here, $D_{\alpha}$ is an anomalous diffusion coefficient of dimensions $\mathrm{m}^{\alpha} \mathrm{s}^{-1}$ and
the regularization parameter $K$ has dimensions $\mathrm{m}^{-1}$. For $\alpha=2$, $\boldsymbol{\zeta}(t)$ reduces to
Gaussian white noise. Using an approach similar to that of Ref.~\cite{2004Bu}, the Fokker-Planck equation corresponding to Eq.~\eqref{e:noise}
is found to be
\begin{eqnarray}
\label{e:frac}
 \frac{\partial}{\partial t} P = 
\left\{D_{\alpha}
\left[  K^{\alpha} -(K^{2} - \nabla^{2})^{\alpha / 2} \right]
+ D_0 \nabla^{2} \right\}P,
\end{eqnarray}
where we also included the contribution from normal diffusion. In Fourier space, the solution of Eq.~\eqref{e:frac} reads
\begin{equation}
\label{e:sol}
\hat{P}(t,\bs{k}) = e^{D_{\alpha}\left[K^{\alpha} -  (K^{2} + |\bs k|^{2})^{\alpha / 2}\right] t - D_0 |\bs k|^2 t}.
\end{equation}
Eq.~\eqref{e:sol} provides a good fit to the the long time simulation data in Figs.~\ref{fig_pos_pdf}~C and \ref{fig_pos_pdf_1D}, especially
in the asymptotic regime. It is worth emphasizing that, although the motion of the tracers at long times is non-Gaussian and described
by a fractional diffusion equation, the asymptotic mean square displacement exhibits normal growth, $\lan \Gd[\bs x(t)]^2 \ran\propto t$.
\par
On intermediate time scales, when the velocity autocorrelations are already decaying, but still non-neglible due to their $t^{-3}$ scaling
(see Fig.~\ref{fig_correlations}), the transient behavior of the position PDF can interpreted as a superposition of two distinct components: i)
Quasi-ballistic tracer displacements, which are remnants of the short time dynamics and may dominate the tails of the tracer position distribution,
and ii) fractional diffusive behavior due to the onset of tracer scattering by multiple swimmers. A quantitative comparison suggests that the measurements
of tracer diffusion in \emph{Chlamydomonas} suspensions by Leptos {\em et al.}~\cite{2009LeEtAl_Gold}, who focused on the range $t\lesssim 2 \ \mathrm{s}$,
are exploring this intermediate regime.

\section{Conclusions}
Understanding the mixing and swimming strategies of algae and bacteria is essential for
deciphering the driving factors behind evolution from unicellular to multicellular
life \cite{2010Drescher_PNAS}. Recent experiments on tracer diffusion in dilute suspensions
of unicellular biflagellate~\emph{Chlamydomonas reinhardtii} algae \cite{2009LeEtAl_Gold} have
shown that microorganisms are able to significantly alter the flow statistics of the surrounding fluid,
which may result in anomalous (non-Gaussian) diffusive transport of nutrients throughout the flow. 
\par
Here, we have developed a systematic theoretical description of anomalous tracer diffusion
in dilute, active suspension. We demonstrated analytically and by means of GPU-based simulations
that, depending on the distance scaling of microflows, qualitatively different flow field
statistics can be expected. For colloidal-type flow fields that scale as $r^{-1}$ (due to the presence
of an external force), the local velocity fluctuations in the fluid are predominantly Gaussian even at
very small volume filling fraction, as expected from the classical central limit theorem. By contrast, flow fields
that rise as $r^{-2}$ or faster in the vicinity of the swimmer will exhibit L\'evy signatures. Very
recent measurements by Rushkin {\em et al.}~\cite{2010Rushkin} appear to confirm this prediction. When the statistics are non-Gaussian,
our results show that the asymptotic convergence properties of velocity fluctuations in active swimmer suspensions are
well-approximated by truncated L\'evy random variables \cite{1994Stanley}.
\par
With regard to experimental measurements, it is important to note that the tracer velocity is a well-defined observable
only if thermal Brownian is negligible (corresponding to the limit $D_0=0$). Otherwise, the associated displacements over a
time-interval $\Gd t$ contain a component that scales as $\sqrt{\Gd t}$. This fact must be taken into account, when one attempts
to reconstruct velocity distributions from discretized trajectories: If thermal Brownian motion is a relevant contribution in the
tracer dynamics, the measured distributions will vary depending on the choice of the discretization interval. 
\par
Our analysis further illustrates that, for homogeneous suspensions, the angular shape of the swimmer flow field
is not important for the local velocity distribution, which is dominated by the radial flow structure. By contrast,
the temporal decay of the velocity correlations sensitively depends on the angular topology of the individual swimmer flow
fields. Specifically, our analytic calculations predicts that velocity autocorrelations in a dipolar
swimmer suspension vanish algebraically as  $t^{-3}$. This prediction could in principle
be tested experimentally by monitoring the flow field at a fixed point in the fluid, using
a setup similar to that in Ref.~\cite{2010Rushkin}.
\par
Finally, we propose that the asymptotic tracer dynamics can be approximated by a fractional
diffusion equation with linearly growing asymptotic mean square displacement.  It would be interesting
to learn whether the fractional evolution of the tracer position distributions at long times can be confirmed
experimentally. This, however, will  require observational time spans that go substantially beyond those considered in
Ref.~\cite{2009LeEtAl_Gold}.
\par
In conclusion, the correct and complete interpretation of experimental data requires an extension of Brownian motion beyond
the currently existing approaches~\cite{1995HaJu,2000MeKl,2004MeKl}. Although many challenging questions remain -- in particular,
regarding the consistent formulation of a generalized diffusion theory that combines L\'evy-type fluctuations with time correlations -- we
hope that the present work provides a first step towards a better understanding of present and future experiments.

\end{document}


\title{Supplementary material}
\author{Irwin M. Zaid}
\email{irwin.zaid@physics.ox.ac.uk}
\author{J\"orn Dunkel}
\email{jorn.dunkel@physics.ox.ac.uk}
\author{Julia M. Yeomans}
\email{j.yeomans1@physics.ox.ac.uk}
\affiliation{Rudolf Peierls Centre for Theoretical Physics, University of Oxford, 1 Keble Road, Oxford OX1 3NP, United Kingdom}
\date{\today}

\begin{abstract}
These supplementary notes summarize analytical calculations for the local velocity distributions,
the velocity autocorrelation functions and the mean square displacement of tracer particles in dilute
suspensions of active swimmers. Section~\ref{s:intro} introduces the model and briefly summarizes a few well-known facts
on characteristic functions required for the subsequent parts. Section~\ref{s:alpha} discusses details and steps that are
essential for the derivation of analytical results for the velocity statistics. To illustrate the general procedure, we will
consider a simplified swimmer model that is helpful for understanding how the distance scaling of effective swimmer flow fields
is reflected in the velocity statistics of tracers. Section~\ref{s:dipol} provides an analogous analysis for a more realistic (regularized)
dipolar swimmer model. Section~\ref{s:numerics} summarizes details of the numerical simulations.
 \end{abstract}

\pacs{
}
\maketitle
\renewcommand{\kBT}{kT}

\section{Summary of model assumptions}
\label{s:intro}

\subsection{Tracer dynamics}

We consider the 3D motion $\bs x(t)=(x_i(t))$ of a passive tracer particle in a fluid that contains $\gs=1,\ldots,N$ active particles
(such as algae or bacteria), which are described by phase space coordinates~$\Gc:=\{\Gc^\gs\}=\{(\bs X^\gs(t),\dot{\bs X}^\gs(t))\}$. We
assume that, in good approximation, the tracer particle does not affect the swimmers, so that $\Gc(t)$ is approximately independent of $\bs x(t)$.
Neglecting Brownian motion effects (corresponding to $D_0=0$ in Eq.~[1] of the main text), the tracer motion can be described by the overdamped equation
(low Reynolds number or Stokes regime) 
\be
\label{e:LE}
\label{e:LE-a}
\dot{\bs x}(t)= \bs u_N(t,\bs x)
\csp
\bs u_N(t, \bs x)=\sum_{\gs=1}^N\bs u(\bs x|\Gc^\gs(t)).
\ee
Here, $\bs u_N$ denotes the velocity field generated by $N$ swimmers and  $\bs u(\bs x|\Gc^\gs)$ the contribution of an individual swimmer~$\gs$ to the fluid
flow at position $\bs x$. In Sections~\ref{s:alpha} and \ref{s:dipol} below, we shall study two different models for $\bs u(\bs x|\Gc^\gs)$. 

\subsection{Swimmer dynamics and statistics}

We restrict our considerations to the  dilute limit of small swimmer concentrations. In this case, we may assume that the swimmers move approximately on straight lines
at constant velocity $\dot{\bs X}^\gs(t)\equiv \bs V^\gs_0$. Then, $\Gc^\gs(t)$ is uniquely determined by its initial condition
$\Gc^\gs_0:=\Gc^\gs(0)=(\bs X_0^\gs,{\bs V}^\gs_0)$, and we have
\be
\bs X^\gs(t)=\bs X_0^\gs+t {\bs V}^\gs_0. 
\ee
Throughout, it is assumed that the swimmer speed is roughly the same and equal to $V$ and that, initially, the swimmers are uniformly distributed in a large spherical  container (radius $\Gl$),  i.e., the initial conditions $\Gc^\gs_0=(\bs X_0^\gs,{\bs V}^\gs_0)$ are i.i.d. random variables with joint probability density function (PDF)
\be\label{e:S_PDF}
\Phi_N(\Gc_0)=\prod_{\gs=1}^N\Phi_1(\Gc^\gs_0)
\csp
\Phi_1(\Gc^\gs_0)=
\f{\mcal{I}(\bs X_0^\gs|\mbb{V})}{(4\pi/3) \Gl^3}\;\f{\gd(|\bs V_0^\gs|-V)}{4\pi V^2},
\ee
where $\mcal{I}(\bs y|\mbb{V})$ is the indicator function of the spherical container volume $\mbb V=\{\bs x\,:\,|\bs x|\le \Gl\}$, 
\be
\mcal{I}(\bs y|\mbb{V}):=
\begin{cases}
1,&   \bs y\in \mbb{V},\\
0, & \bs y\not\in \mbb{V}.
\end{cases}
\ee
We are interested the velocity distribution at the center of the volume, formally defined by
\bse
\be\label{e:def_vPDF}
\phi_{N,\Gl}(\bs v) &:=&\lan \gd(\bs v-\bs u_N(0,\bs 0))\ran,
\ee
and, moreover, in the flow field autocorrelation function at a fixed point in the fluid far from the boundaries,
\be\label{e:def_CF}
C_{N,\Gl}(t,)
:=\lan \bs u_N(t,\bs 0) \bs  u_N(0,\bs 0)\ran,
\ee 
\ese
where $\lan\,\cdot\,\ran $ indicates an average with respect to the initial swimmer distribution~\eqref{e:S_PDF}. Generally,
we are interested in evaluating these quantities in some suitably defined thermodynamic limit $N,\Gl\to \infty$.

\subsection{Characteristic function}

Consider the characteristic function $\chi_\phi$ of some velocity PDF $\phi(\bs v)$, defined by 
\be
\chi_\phi(\bs q)
:=
\int\diff^d v \;
\exp(-i\bs q \bs v)\;\phi(\bs v).
\ee
Then, by inverse Fourier transformation, we have
\be
\phi(\bs v)=\f{1}{(2\pi)^d}\int\diff^d q \;
\exp(i\bs q \bs v)\;\chi_\phi(\bs q).
\ee
Given $\chi_\phi(\bs q)$,  the second and fourth velocity moment can be obtained by differentation
\be
\lan  |\bs v|^2\ran_\phi=-\triangle_{\bs q} \chi_\phi(\bs 0)
\csp
\lan  |\bs v|^4\ran_\phi= \triangle_{\bs q}^2 \chi_\phi(\bs 0).
\ee
In particular, for 3D spherically symmetric velocity PDF   $\phi(\bs v)=\tilde \phi(v)$ with $v=|\bs v|$, we have $\chi_\phi(\bs q)=\tilde{\chi}_\phi(q)$, where $q=|\bs q|$. In this case, the differential operators  simplify to
\bse
\be
\triangle_{\bs q}
=
\left(\p_q^2 +\f{2}{q}\p_q\right)
\csp
\triangle^2_{\bs q}
=
\left(\p_q^2 +\f{2}{q}\p_q\right)^2=\p_q^4  +\f{4}{q}\p^3_q .
\ee
\ese
\par
Below, we shall show that the velocity PDF of a tracer in the presence of  active swimmers can be approximated
by a tempered L\'evy stable distribution with characteristic function
\be \label{e:tlevy}
\tilde{\chi}_\phi(q)=\exp\{-[(c\; q^2+\mu^2 )^{\ga/2}-\mu^\ga]\},
\ee
which yields for the second and fourth velocity moment
\bse\label{e:ansatz_irwin_moments}
\be
\lan  |\bs v|^2\ran
&=&
3 \ga c\;   \mu ^{\ga-2},
\\
\lan  |\bs v|^4\ran
&=&
15 \ga c^2  \mu ^{\ga-4} \left[2+\ga  \left(\mu^{\ga }-1\right)\right].
\ee
\ese
The tempered L\'evy distribution in Eq.~\eqref{e:tlevy} reduces to either the Gaussian prediction
from the central limit theorem as $\mu$ becomes large (many swimmers) or the L\'evy stable distribution
from the generalized central limit theorem \cite{1954GnKo} as $\mu \to 0$ (unregularized swimmers).

\section{First example: Regularized co-oriented model}
\label{s:alpha}

We first consider the model interaction
\bse\label{e:alpha_model}
\be
 \bs u(\bs x| \Gc^\gs_0)
= (\gk \bs V_0^\gs)\;\f{\eps^n}{|\bs r^\gs(t)|^n+\gl^n},
\ee
where $\eps$ can be interpreted as the swimmer radius,  $0<\gk < 1$ is a dimensionless coupling constant, $\gl$ a regularization parameter, $\bs V^\gs_0$ the swimmer velocity (assumed to be constant),  and
\be
\bs r^\gs(t):=\bs X^\gs(t)-\bs x,
\ee
\ese
is the vector connecting the swimmer position at time $t$, $\bs X^\gs(t)=\bs X^\gs_0+ t\bs V^\gs_0$, with the space point $\bs x$.

\subsection{Velocity PDF of the tracer particle}

Inserting the Fourier representation of the $\gd$-function, we can rewrite \eqref{e:def_vPDF} as
\be
\phi_{N,\Gl}(\bs v) :=
\f{1}{(2\pi)^3 }\int d^3q \int \diff \Gc_0\, \Phi_N(\Gc_0)\,
\exp\left\{i\bs q\left[\bs v-  \sum_{\gr}\bs u(\bs 0| \Gc^\gr_0)\right]\right\}.
\ee
Using spherical variables,  $\bs X_0^\gs = R\bs \Go$ and $\bs V_0^\gs = V\hat{\bs \Go}$ for both the initial swimmer positions and velocities, we find
\be
\phi_{N,\Gl}(\bs v)
=
\left[\f{(4\pi/3)^{-1}}{\Gl^3}\right]^N 
\f{1}{(2\pi)^{3}}
\int \diff^3q \exp(i\bs q \bs v )\;
\left[
\int_0^\Gl  \diff R\; R^2
\int\diff^{2}\hat{\Go}\; 
\exp\biggl(-i  \;\gk V\,\f{\eps^n}{{R}^n+\gl^n}  \bs q \hat{\bs \Go}\biggr)
\right]^N.
\ee
Introducing a rescaled radial position variable $y=R/\Gl$, and performing the angular integral over $\diff^{2}\hat{\Go}$,  we obtain
\bse
\be
\phi_{N,\Gl}(\bs v)
=
\f{1}{(2\pi)^{3}}
\int \diff^3q \exp(i\bs q \bs v )\;K^N_n(\bs q),
\ee
where
\be
K^N_n(\bs q)
&=&
\left[
3\int_0^1  \diff y\; y^2\;
\f{\sin\left[A_n(y)|\bs q|\right]}{A_n(y)  |\bs q|}
\right]^N
\label{e:def_K_alpha}
\ee
is the characteristic function of $\phi_{N,\Gl}$ with
\be
A_n(y):=\f{ (\eps/\Gl)^{n}}{{y}^n+(\gl/\Gl)^{n}} \; \gk V.
\ee
\ese
It is useful to rewrite $K^N_n$ in the equivalent form
\be\label{e:KN_integral}
K^N_n(\bs q)
&=&
\left[
3\int_0^1\diff \xi \int_0^1 \diff y\; y^2\;
\cos(A_n \,|\bs q|\;\xi)
\right]^N
=:[K_n(q)]^N,
\ee
where $K_n(q)$ is the characteristic function for the one-swimmer case.

\subsubsection{Exact second and fourth velocity moments}
We note that,  for $m\in\mbb{N}$,
\bse
\be
K_n(0)&=&1,
\\
\p_q^{2m} K_n(0)
&=&
(-1)^m\; \f{3}{2m+1}\int_0^1 \diff y\; y^2\; A^{2m} ,
\\
\p_q^{(2m-1)} K_n(0)
&=&0,
\\
\lim_{q\to 0}
\f{1}{q}\p_q^{(2m-1)} K_n(q)
&=&
(-1)^m \f{3}{2m+1} \int_0^1 \diff y\; y^2\;
 A^{2m}
 \;=\;
\p_q^{2m} K_n(0).
 \ee
 \ese
From these expressions, one finds
\bse
\be
\p_q K_n^N(\bs 0)
&=&
0,\\
\p_q^2K^N_n(\bs 0)
&=&
N  [\p_q^2 K_n(0)],
\\
\p_q^3 K^N_n(\bs 0)
&=&
0,\\
\p_q^4 K^N_n(\bs 0)
&=&
3(N-1) N  [\p_q^2 K_n(0)]^2+
N  [\p_q^4 K_n(0)],
\ee
\ese
and, therefore,
\bse\label{e:moments_char_N}
\be
\triangle K_n^N(0)
&=&
3N [\p^2_q K_n(0)],
\\
\triangle^2 K_n^N(0)
&=&
15 N(N-1)  [\p_q^2K_n(0)]^2+5N \p^4_q K_n(0).
\ee
\ese
Equations~\eqref{e:moments_char_N} hold for any spherically symmetric distribution in 3D. 
\par
From the above expressions, one finds the following exact formulas for the velocity moments in this model:
\bse\label{e:alpha_moments}
\be
\lan  |\bs v|^2\ran_\phi
&=&
3N \int_0^1 \diff y\; y^2\,
\left[A_n(y)\right]^2,
\\
\lan  |\bs v|^4\ran_\phi
&=&
\f{5 (N-1)}{3N} \left[\lan  |\bs v|^2\ran_\phi \right]^2+
3 N  \int_0^1 \diff y\; y^2\,[A_n(y)]^{4},
\ee
where, using the abbreviation $\ell:=\gl/\Gl$,
\be
\int_0^1 \diff y\; y^2\,[A_n(y)]^{2}&=&
\f{(\gk V)^2}{n} 
\left(\frac{\epsilon }{\gl }\right)^{2 n } 
\left\{\frac{\ell ^{n }}{1+\ell^{n }}+
\frac{1}{3} (n-3) \text{HypGeom2F1}\left[1,\frac{3}{n },\frac{3+n }{n },-\ell^{-n }\right]\right\},
\qquad
\\
\int_0^1 \diff y\; y^2\,[A_n(y)]^{4}&=&
\frac{(\gk V)^4}{6 n ^3}
\left(\frac{\epsilon }{\gl }\right)^{4 n } 
\quad
\biggl\{
\frac{3(n-1 )}{\left(1+ \ell^{n }\right)^3}  
\biggl[
(2 n-3 )\, \ell^{n} +
  (5 n -6)\,\ell^{2n }+
\f{9-18n+11 n^2  }{3(n-1)}\,\ell^{3 n }
\biggr]+
\notag\\&&
\qquad\qquad\qquad\qquad
 (n-1)(n-3)  (2n-3) 
\text{HypGeom2F1}\left[1,\frac{3}{n },\frac{3+n }{n },-\ell^{-n }\right]
\biggr\}.
\ee
\ese
For example, for the ($n=2$) model, these expressions simplify to
\bse
\be
\int_0^1 \diff y\; y^2\,[A_n(y)]^{2}
&=&
\f{1}{2}\left(\f{\eps}{\Gl}\right)^4
\left[
\left(\f{1}{\ell}\right)
  \arctan\left(\f{1}{\ell}\right)
-
\frac{1} {1 +\ell^2}
\right],
\\
\int_0^1 \diff y\; y^2\,[A_n(y)]^{4}
&=&
\frac{1}{16}\left(\frac{\epsilon }{\Lambda }\right)^4\left(\frac{\epsilon }{\lambda }\right)^4
\left\{
\left(\frac{1}{\ell}\right) \arctan\left(\frac{1}{\ell}\right)+
\frac{1-\ell^2}{\left(1+\ell^2\right)^2}+
\frac{8}{3}\frac{\ell^2}{\left(1+\ell^2\right)^3}
\right\}.
\ee
\ese
The first integral, and hence $\lan  |\bs v|^2\ran_\phi$, diverges as $1/\gl$ if one lets the cut-off $\gl\to 0$ (which is equivalent to $\ell\to 0$).
More precisely, in this limit, Eqs.~\eqref{e:alpha_moments} reduce in leading order of $(\Gl/\gl)$ to
\bse\label{e:alpha2_moments_approx}
\be
\lan |\bs v|^2\ran_\phi
&\simeq&
\f{3\pi}{4}(\gk V)^2 \;\gvf
\left(\f{\eps}{\gl}\right),
\\
\lan |\bs v|^4\ran_\phi
&\simeq&
\f{3\pi}{32}(\gk V)^4 \;\gvf
\left(\f{\eps}{\gl}\right)^5,
\ee
\ese
where $\gvf:=N(\eps/\Gl)^3$ is the volume filling fraction.

\subsubsection{Approximation by a tempered L\'evy distribution}

We would like to approximate the exact characteristic function $K^N_n$ from Eq.~\eqref{e:KN_integral} by the tempered L\'evy law
\be\label{e:ansatz_irwin}
\tilde{\chi}_\phi(q)=\exp\{-[(c\; q^2+\mu^2 )^{\ga/2}-\mu^\ga]\},
\ee
which exhibits quasi-Gaussian behavior for small $q$, corresponding to large velocities, 
\be\label{e:ansatz_irwin_gauss}
\tilde{\chi}_\phi(q)\simeq
\exp\left( -\ga  \mu ^{\ga-2} c \; \f{q^2}{2}\right).
\ee
To motivate the ansatz~\eqref{e:ansatz_irwin}, we write~\eqref{e:def_K_alpha} as
\be
K^N_n(\bs q) =\exp\left\{- N\ln[K_n(q)]\right\},
\ee
and consider the limit $\gl=0$. In this case, the double integral for the one-swimmer characteristic function
$K_n(q)$ from  Eq.~\eqref{e:KN_integral} can be calculated exactly for $n=1,2,3$ in terms of trigonometric, hypergeometric,
and sine integral functions. By expanding the resulting expressions $\ln[K_n(q)]$ for a large volume $\Gl \gg \eps$, one obtains
for $n=1$ a Gaussian limiting distribution
\bse\label{e:asymp}
\be\label{e:alpha1_gauss}
K_1^N(\bs q)
\simeq
\exp\left(- T_1\,q^2 \right)
\csp
T_1= \f{1}{2}N \left(\f{\eps}{\Gl}\right)^2 (\gk V)^2.
\ee
By contrast,  for $n\ge 2$  the limiting distribution is found to be of the Holtsmark type, i.e.,
\be\label{e:ansatz_irwin_K}
K^N_n(\bs q)
&\simeq&
\exp\left(- T_n\, q^{3/n} \right)
\csp 
T_n=t_n\, N \left(\f{\eps}{\Gl}\right)^3 (\gk V)^{3/n} 
\ee
where $t_n$ is a constant of order unity; specifically, $t_2={\sqrt{8 \pi } }/{5}$ and $t_3=\pi/4$. For comparison, if we let $\mu\to 0$ in Eq.~\eqref{e:ansatz_irwin}, we obtain
\be\label{e:irwin_asymp}
\tilde{\chi}_\phi(q)=\exp\left(- c^{\ga/2}\; q^\ga\right),
\ee
\ese
Thus, by comparing  with~\eqref{e:asymp} and \eqref{e:irwin_asymp},  we can identify
\bse\label{e:beta}
\be
n=1\quad &\Rightarrow&\quad  \ga=2, \\
n\ge 2\quad &\Rightarrow&\quad  \ga=\f{3}{n}.
\ee
\ese
Before determining the remaining parameters $(c,\mu)$, it is worthwhile to note that the effective temperature scales differently with volume and swimmer number for $n=1$ and $n\ge 2$, respectively: In the case of a colloidal-type velocity field with $n=1$, the effective temperature $T_1$ is proportional to the \emph{area} filling fraction $N(\eps/\Gl)^2$, see Eq.~\eqref{e:alpha1_gauss}, whereas for swimmer-type flow fields with $n\ge 2$ the effective temperature $T_n$  scales with the \emph{volume} filling fraction 
\be
\gvf:=N(\eps/\Gl)^3,
\ee 
see Eq.~\eqref{e:ansatz_irwin_K}. This suggest that, for $n=1$, the appropriate thermodynamic limit corresponds to $N,\Gl\to \infty $ while keeping the ratio $N/\Gl^2$ fixed, whereas for $n\ge 2$  one should let $N,\Gl\to \infty $ such that $N/\Gl^3$ remains constant.
\par
It remains to discuss how to identify the parameters $(c,\mu)$, which we shall do next for the cases $n=1,2,3$. For $n=1$ the procedure is  rather straightforward;  in the case of  $n=2,3$, we are going to determine $(c,\mu)$ by matching the second and fourth velocity moments of the tempered L\'evy ansatz~\eqref{e:ansatz_irwin}, which were given in Eq.~\eqref{e:ansatz_irwin_moments}, to the exact moments~\eqref{e:alpha_moments}.

\paragraph*{$n=1$.--}
In this case, according to Eq.~\eqref{e:beta}, we have $\ga=2$, and the ansatz~\eqref{e:ansatz_irwin} reduces to the Gaussian 
\bse
\be
\tilde{\chi}_\phi(q)=\exp(-c\; q^2).
\ee
By comparing with \eqref{e:alpha1_gauss}, we see the $c=T_1$, i.e.,
\be
c=\f{1}{2} N \left(\f{\eps}{\Gl}\right)^2\ (\gk V)^2\;\left[1+\ell+2\ell \log(\ell)\right],
\ee
\ese
where we now also included the leading order corrections in $\ell=\gl/\Gl$.

\paragraph*{$n=2$.--}
In this case, we have $\ga=3/2$ and the velocity moments~\eqref{e:ansatz_irwin_moments} of the tempered L\'evy ansatz~\eqref{e:ansatz_irwin} take the form
\bse\label{e:ansatz_irwin_moments_2}
\be
\lan  |\bs v|^2\ran
&=&
\f{9}{2} c\;   \mu ^{-1/2},
\\
\lan  |\bs v|^4\ran
&=&
\f{45}{2} c^2  \mu ^{-5/2} \left[2+\f{3}{2}  \left(\mu^{3/2}-1\right)\right].
\ee
Solving these equations for $(c,\mu)$ yields
\be
c
=
\f{2}{9} \left(\frac{5}{3}\right)^{1/3}
\frac{ \lan |\bs v|^2\ran^{5/3}}{\left(3 \lan |\bs v|^4\ran-5 \lan |\bs v|^2\ran^2\right)^{1/3} }
\csp
\mu
=
\left(\frac{5}{3}\right)^{2/3}\frac{ \lan |\bs v|^2\ran^{4/3}}{\left(3 \lan |\bs v|^4\ran-5 \lan |\bs v|^2\ran^2\right)^{2/3}}.
\ee
Here, we can inserting for $\lan |\bs v|^2\ran$ and $\lan |\bs v|^4\ran$ the exact expressions~\eqref{e:alpha_moments}; this gives the fit curves shown in Fig. 2 of the main paper. Furthermore, by expanding the resulting formula for large volume  $\Gl\gg\eps,\gl$, we find
\be\label{e:mu_2}
c
=
\f{1}{3} \left(\frac{5\pi^4}{12}\right)^{1/3}
\left(\kappa V\right)^2
\gvf^{4/3}
\csp
\mu
=
\left(\frac{10 \pi }{3}\right)^{2/3} \left(\f{\gl}{\eps}\right)^2 \gvf^{2/3}.
\ee
\ese

\paragraph*{$n=3$.--}
In this case, we have $\ga=1$ and the general expression for the velocity moments of the tempered L\'evy law from Eq.~\eqref{e:ansatz_irwin_moments} reduce to 
\bse\label{e:ansatz_irwin_moments_3}
\be
\lan  |\bs v|^2\ran
&=&
3 c\;   \mu ^{-1},
\\
\lan  |\bs v|^4\ran
&=&
15  c^2  \mu ^{-3} \left(\mu +1\right).
\ee
Solving these equations for $(c,\mu)$ yields
\be
c=\frac{5 \lan |\bs v|^2\ran^3}{9 \lan |\bs v|^4\ran-15 \lan |\bs v|^2\ran^2}
\csp
\mu=\frac{5 \lan |\bs v|^2\ran^2}{3 \lan |\bs v|^4\ran-5 \lan |\bs v|^2\ran^2}.
\ee
Inserting for $\lan |\bs v|^2\ran$ and $\lan |\bs v|^4\ran$ the exact expressions~\eqref{e:alpha_moments} and expanding for $\Gl\gg\eps,\gl$, we obtain
\be\label{e:mu_3}
c
=
\f{5}{3} 
\left(\kappa V\right)^2
 \gvf^2
\csp
\mu
=
5 \left(\f{\gl}{\eps}\right)^3 \gvf.
\ee
\ese
\par
Note that for a vanishing $\gl\to 0$, the parameter $\mu$ goes to zero in Eqs.~\eqref{e:mu_2} and~\eqref{e:mu_3}, which implies that the second moment $\lan  |\bs v|^2\ran$ diverges. This also illustrates why for $n \ge 2$  -- or,  more precisely, for $n\ge 3/2$ if one allows for non-integer exponents $n$ -- there is no convergence to a Gaussian distribution in the limit $\gl\to 0$ .

\subsection{Velocity autocorrelation function}

We are interested in the fluid's velocity autocorrelation function~\eqref{e:def_CF}  for the power-law model~\eqref{e:alpha_model}. We
start from Eq.~\eqref{e:def_CF}, which can be written as
\be
C_{N,\Gl}(t)
&=&
\sum_{\gs,\gr}
\left\lan 
u_i(\bs x| \Gc^\gs(t))\, u_i(0| \Gc^\gr_0)
\right\ran.
\ee
Assuming, as before, that the initial swimmer position and velocities are distributed according to \eqref{e:S_PDF}, we can simplify
\be
C_{N,\Gl}(t)
&=&
N (\gk V)^2\,
\left\lan 
 \;\f{\eps^{n}}{(|\bs X_0^\gs + t \bs V_0^\gs|^n+\gl^n)}
 \;\f{\eps^{n}}{ (|\bs X_0^\gs|^n+\gl^n)}
 \right\ran
 =:
 N (\gk V)^2\,
 c_n(t).
\ee
Using spherical velocity and position  variables,  $\bs X_0^\gs = X\bs \Go$ and $\bs V_0^\gs = V\hat{\bs \Go}$, and inserting the one-swimmer PDF from Eq.~\eqref{e:S_PDF},  we find
 \be
 c_n(t)  &=&
\int\diff \Go\int_0^\Gl\diff X \;X^2\;\f{3}{4\pi \Gl^3}
\int\f{\diff \hat\Go }{4\pi}
 \;\f{\eps^{n}}{(|X\bs \Go + t V\hat{\bs\Go}|^n+\gl^n)}
 \;\f{\eps^{n}}{ (X^n+\gl^n)}.
 \label{e:alpha_vacf_3}
\ee
Introducing rescaled  variables
\bse\label{e:rescaling}
\be
y:=X/\Gl
\csp
s:=t V/\Gl
\csp
\ell:=\gl/\Gl
\csp 
z:=\bs \Go \hat{\bs\Go}
\ee
and noting that
\be
|X \bs \Go + t V\hat{\bs\Go}|
=
\Gl \left[
y^2 + s^2 +2syz)
\right]^{1/2},
\ee
\ese
we can rewrite~\eqref{e:alpha_vacf_3} in the form
\be\label{e:c_alpha}
c_n(t)
=
\left(\f{\eps}{\Gl}\right)^{2n}
\int\diff \Go\int_0^1\diff y \;y^2\;\f{3}{4\pi}
\int\f{\diff \hat\Go }{4\pi}
 \;\f{1}{(y^2 + s^2 +2syz)^{n/2}+\ell^n}\;
 \f{1}{ y^n+\ell^n}.
\ee
By virtue of  the identity
 \be\label{e:identity}
 \int \diff {\Go} \int \diff \hat{\Go}\; f(z)=(4\pi)(2\pi)\int_{-1}^1 
\diff z\;f(z),
 \ee
Eq.~\eqref{e:c_alpha} can expressed as
\bse
\be\label{e:c_double}
c_n(t)
&=&
\f{3}{2}\left(\f{\eps}{\Gl}\right)^{2n}
\int_0^1\diff y \;\f{y^2}{ y^n+\ell^n}\;
J_n(s,y),
\ee
where
\be
J_n(s,y)
&:=&\int_{-1}^1\diff z
\;\f{1}{(y^2 + s^2 +2syz)^{n/2}+\ell^n}\;.
\ee
\ese
We next provide explicit results for $n=1$ and $n=2$.

\paragraph*{$n=1$.--} In this case, we find
\bse
\be
J_1(s,y)
&=&
\frac{s+y-|s-y|}{s y}+\f{\ell}{s y}  \log\left[\frac{\ell + |s-y|}{s+y+\ell }\right].
\ee
The remaining integral $y$-integration in Eq.~\eqref{e:c_double} can be easily computed numerically. However, the correlation function $c_1(t)$ can also be  expressed analytically in terms of  the polylogarithm $\mrm{Li}_q(z)$, defined by 
\be
\mrm{Li}_q(z):= \sum_{k=1}^\infty\f{z^k}{k^q}
\csp
 |z| <1
\ee
and analytic continuation for $|z|\ge 1$.  $\mrm{Li}_q(z)$ is real valued for real $z\le 1$ and possesses the 
integral representation
\be
\mrm{Li}_q(z)=\f{z}{\Gc(q)}\int_0^\infty\diff k\;\f{k^{q-1}}{e^k-z}.
\ee
Considering a sufficiently large volume such that  $0\le s\le1$, we obtain
\be
c_1(t)
&=&\notag
-\f{3}{2}\left(\f{\eps}{\Gl}\right)^{2}\frac{1}{s} \;\Re
\biggl\{
s^2+ (2 \ell-2) s+  2 \ell (1+\ell ) \mrm{arctanh}\left(\frac{s}{1+\ell }\right)+
\\&&\quad
\ell^2 \log\left(\frac{\ell }{s+\ell }\right)\left[4+ \log\left(\frac{s}{s+2 \ell }\right)
 \right]+
 \ell s \log\left[\frac{(1+\ell )^4 }{(s+\ell )^4}-\frac{(1+\ell )^2 s^2}{(s+\ell )^4}\right]
 +
\\&&\notag\quad
\ell ^2 \left[
\mrm{Li}_2\left(\frac{s+\ell }{s}\right) -
\mrm{Li}_2\left(\frac{1+\ell }{s}\right) -
\mrm{Li}_2\left(-\frac{\ell }{s}\right)  +
\mrm{Li}_2\left(-\frac{1+\ell }{s}\right) +
\mrm{Li}_2\left(\frac{\ell }{s+2 \ell }\right) -
\mrm{Li}_2\left(\frac{s+\ell }{s+2 \ell }\right)
\right]
\biggr\},
\ee
with $\Re$ denoting the real part. In particular, in the thermodynamic limit $N,\Gl\to\infty$ such that $\gvf=N(\eps/\Gl)^2=const.$, we find that the autocorrelation function $C_{N,\Gl}(t)=N(\gk V)^2c_n(t)$ becomes constant
\be
C^{n=1}_{\infty}(t)=3\; \varphi \,(\kappa V)^2. 
\ee
\ese
This situation, however, is unrealistic for real swimmers, which typically generate flow fields that decay with $n\ge 2$.
\paragraph*{$n=2$.--} In this case, we find
\bse
\be
J_2(s,y)
=
\frac{1}{2 s y}\log\left[1+\frac{4 s y}{(s-y)^2+\ell ^2}\right].
\ee
Considering again $0\le s\le1$, the correlation function $c_2(t)$ may be written in terms of the Dilogarithm $\mrm{Li}_2(z)$ as
\be
c_2(t)
&=&\notag
-\f{3}{4}\left(\f{\eps}{\Gl}\right)^{4} \frac{1}{s}\;\Re
\biggl\{\f{1}{2}
\log(s)\log\left[1+\frac{4 s}{(s-1)^2+\ell ^2}\right]+
\log\left(\frac{s-1+i \ell }{s+1+i \ell }\right)
\log\left(\frac{1+\ell ^2}{s+2 i \ell }\right)
+
\\&&\qquad\qquad
\mrm{Li}_2\left(\frac{s-1-i \ell }{s}\right) -
\mrm{Li}_2\left(\frac{s+1-i \ell }{s}\right)  + 
\mrm{Li}_2\left(\frac{s-1-i \ell }{s-2 i \ell }\right) -
\mrm{Li}_2\left(\frac{s+1-i \ell }{s-2 i \ell }\right) 
\biggr\},
\ee
with $\Re$ denoting the real part.
In the thermodynamic limit $N,\Gl\to\infty$ such that $\gvf=N(\eps/\Gl)^3=const.$, we find at large times $t\gg \tau_\eps:=\epsilon/V$ for the full autocorrelation function
\be\label{e:alpha=2_C_longtime}
C^{n=2}_{\infty}(t)\simeq\frac{3 \pi ^2}{4}  \varphi \,(\kappa V)^2\,  \,\left(\f{\tau_\eps}{t}\right).
\ee
\ese


\section{Second example: Regularized dipolar swimmer model}
\label{s:dipol}

Let us now consider a more realistic dipolar swimmer flow field model, defined by
\bse\label{e:dipol_model}
\be
 \bs v(\bs x| \Gc^\gs_0)
= 
(\gk V) \left(\f{  \eps^2}{|\bs R^\gs|^2+\gl^2}\right)
\left[
3\left(\hat{\bs \Go}^\gs \hat{\bs R}^\gs\right)^2-1
\right]
\hat{\bs R}^\gs 
\csp
\gl\ge 0,
\ee
where $\gk>0$ ($\gk<0$) correspond to an extensile (contractile) swimmer, and
\be
{\bs R}^\gs(t):=\bs X^\gs(t)-\bs x
\csp
\hat{\bs R}^\gs(t):=\f{\hat{\bs R}^\gs(t)}{|\hat{\bs R}^\gs(t)|}
\csp
\hat{\bs \Go}^\gs(t):=\f{\bs V^\gs(t)}{|{\bs V}^\gs(t)|}\equiv \hat{\bs \Go}^\gs_0.
\ee
\ese
As before, we assume, for simplicity, that a swimmer's orientation does not change over time. The dipolar swimmer model~\eqref{e:dipol_model} exhibits the same distance scaling as the
regularized power-law model from Eq.~\eqref{e:alpha_model} with $n=2$, but the directional dependence is different. As a consequence, 
as we shall see below, the velocity PDFs of the two models are very similar but the correlation functions show qualitatively different behavior.

\subsection{Tracer velocity PDF}

\subsubsection{Characteristic function}

Using spherical variables,  $\bs X_0^\gs = X\bs \Go$ and $\bs V_0^\gs = V\hat{\bs \Go}$ for both the initial swimmer positions and velocities, the characteristic function of the velocity PDF can be written as
\be
K_\mrm{S}^N(\bs q)=[K_\mrm{S}(q)]^N
\ee
where the one-swimmer characteristic function is given by
\be
K_\mrm{S}(q)
=
\f{(4\pi/3)^{-1}}{\Gl^3}\f{1}{4\pi}
\int_0^\Gl  \diff R\; R^2
\int\diff^{2}{\Go}\; 
\int\diff^{2}\hat{\Go}\; 
\exp\left\{
-i  (\gk V) \left(\f{  \eps^2}{X^2+\gl^2}\right)
\left[
3\left(\hat{\bs \Go} \bs \Go\right)^2-1
\right]
(\bs q \bs \Go)
\right\}.
\ee
Hence, with $y:=R/\Gl$, $\ell:=\gl/\Gl$ and $z:=\hat{\bs \Go} \bs \Go$,
\be
K_\mrm{S}(q)
=
\f{3}{8\pi}
\int_0^1  \diff y\; y^2
\int\diff^{2}{\Go}\; 
\int_{-1}^1\diff z\; 
\exp\left\{
-i  (\gk V)\left(\f{\eps}{\gl}\right)^2 \left(\f{\ell^2}{y^2+\ell^2}\right)
\left(
3z^2-1
\right)
(\bs q \bs \Go)
\right\}.
\ee
Choosing $\bs q=(0,0,|\bs q|)=(0,0,q)$, with no loss of generality, we find
\bse\label{e:dipol_vel_char}
\be
K_\mrm{S}(q)
&=&
\f{3}{2}
\int_0^1  \diff y\; y^2
\int_{-1}^1\diff z 
\int_0^1\diff\xi\;
\cos(Dq\xi),
\ee
where
\be
D(y,z):=
(\gk V)\left(\f{\eps}{\gl}\right)^2 \left(\f{\ell^2}{y^2+\ell^2}\right)
\left(
3z^2-1
\right).
\ee
\ese
This result can be used to evaluate analytically the moments of the tracer velocity PDF.

\subsubsection{Velocity moments}

From Eqs.~\eqref{e:dipol_vel_char} and~\eqref{e:moments_char_N}, we find the following exact results for the second  and fourth velocity moments:
\bse\label{e:dipol_moments}
\be
\lan |\bs v|^2 \ran
&=&\label{e:dipol_moments_2}
N (\gk V)^2\; 
\left(\f{\eps}{\Gl}\right)^4
\left\{
\f{6}{5}
\left[
\f{\mrm{\arctan}\left({1}/{\ell}\right)}{\ell}-\f{1}{1+\ell^2}
\right]\right\},
\\
\lan |\bs v|^4 \ran
&=&
N (\gk V)^4\; 
\left(\f{\eps}{\Gl}\right)^8
\left\{\f{3}{35}
\left[
\frac{3\,\text{arctan}(1/\ell )}{\ell ^5}
+
\frac{3+8 \ell ^2-3 \ell ^4}{\ell ^4 \left(1+\ell ^2\right)^3}
\right]
\right\}
+\f{5}{3}\f{N-1}{N}\lan |\bs v|^2 \ran^2,
\ee
\ese
where $\ell:=\gl/\Gl$ is the rescaled regularization cut-off. In the small cutoff limit  $\ell\to 0$, we find
\bse\label{e:dipol_moments_approx}
\be
\lan |\bs v|^2\ran
&\simeq&
\f{3\pi}{5}(\gk V)^2 \;\gvf
\left(\f{\eps}{\gl}\right),
\\
\lan |\bs v|^4\ran
&\simeq&
\f{9\pi}{70}(\gk V)^4 \;\gvf
\left(\f{\eps}{\gl}\right)^5.
\ee
\ese
These expressions are quite similar to those obtained for the power-law model with $n=2$, see Eq.~\eqref{e:alpha2_moments_approx}.
The moments \eqref{e:dipol_moments_approx} can be used to determine the parameters of the corresponding tempered L\'evy
velocity distribution~\eqref{e:ansatz_irwin} by means of Eqs.~\eqref{e:mu_2}. However, as we shall see below, the two models give
rise to very different velocity correlations.

\subsection{Velocity autocorrelation function}

We are interested in the velocity autocorrelation function~\eqref{e:def_CF} of the fluid near the center of the volume, which can be written as
\bse
\be
C_\mrm{S}(t)
=
N (\gk V)^2\; 
\left\lan 
\left(\f{  \eps^2}{| {\bs R}(t)|^2+\gl^2}\right)
\left(\f{  \eps^2}{| {\bs R}(0)|^2+\gl^2}\right)
\left\{
3\left[\hat{\bs \Go} \hat{\bs R}(t)\right]^2-1
\right\}
\left\{
3\left[\hat{\bs \Go} \hat{\bs R}(0)\right]^2-1
\right\}
\hat{\bs R}(t) \hat{\bs R}(0) 
 \right\ran.
 \ee
For linear swimmer motions, we further have
 \be
 {\bs R}(t):=\bs X + t \bs V
 \csp
 \bs X=X {\bs \Go}
\csp
 \bs V=V\hat{\bs \Go}
 \ee
 \ese
 with $\bs X$ and $\bs V$  denoting the initial swimmer position and velocity; $\bs \Go$ and  $\hat{\bs \Go}$ are the corresponding unit vectors. Defining the dimensionless velocity autocorrelation function $c_\mrm{S}(t)$ by $C_\mrm{S}(t)= N (\gk V)^2\; c_\mrm{S}(t)$,  we obtain
 \be\label{e:c_dip}
c_\mrm{S}(t)
=
\left\lan 
\left(\f{  \eps^2}{| \bs X +  t V{\bs \Go}|^2+\gl^2}\right)
\left(\f{  \eps^2}{| {\bs X}|^2+\gl^2}\right)
\left[
3\left(\f{X \bs \Go\hat{\bs \Go}+ t V  }{|\bs X + t V \bs \Go|}\right)^2-1
\right]
\left[
3\left({\bs \Go} \hat{\bs \Go}\right)^2-1
\right]
\;
\f{X + t V \bs \Go\hat{\bs \Go}}{|\bs X + t V \bs \Go|} 
 \right\ran.
\ee
In the limit $t=0$, one recovers the second velocity moment, $c_\mrm{S}(0)=\lan |\bs v^2|\ran/[N(\gk V)^2]$. 
\par
For $t>0$, using the notation from Eqs.~\eqref{e:alpha_vacf_3} and~\eqref{e:rescaling}, we have
 \be
c_\mrm{S}(t) =\eps^4
\left\lan 
\left(\f{ 1}{X^2+\gl^2}\right)
\left(
\f{3z^2-1}{X^2+t^2V^2 +2 t V X z+\gl^2}
\right)
\left[
\f{3(Xz+t V )^2 }{X^2+t^2V^2 +2 t V X z}-1
\right]\;
\f{X +t V z}{\sqrt{X^2+t^2V^2 +2 t V X z}} 
 \right\ran
\ee
Substituting $y:=X/\Gl$, $\ell=\gl/\Gl$, $s:=t V/\Gl$ and using the identity~\eqref{e:identity},  this can be written as
\bse\label{e:s_cf_split}
 \be
c_\mrm{S}(t)=
\f{3}{2} \left(\f{\eps}{\Gl}\right)^4\Gd I(t)
\csp
\Gd I(t):=I_1(t)-I_2(t),
\ee
where 
\be
I_1(t) &=&
\int_0^1\diff y
\int_{-1}^1\diff z
\left(\f{y^2}{\ell^2+y^2}\right)
\left(
\f{ 3z^2-1}{\ell^2+y^2+ s^2 +2 s y z}
\right)
\left[
\f{y +s z}{(y^2+ s^2 +2 s y z)^{1/2}} 
\right]
\left[
\f{3(yz+ s)^2}{(y^2+ s^2 +2 s y z)^{2/2}}
\right],
\label{e:s_cf_split_A}
\\
I_2(t) &=&
\int_0^1\diff y
\int_{-1}^1\diff z
\left(\f{y^2}{\ell^2+y^2}\right)
\left(
\f{ 3z^2-1}{\ell^2+y^2+ s^2 +2 s y z}
\right)
\left[
\f{y + s z}{(y^2+ s^2 +2 s y z)^{1/2}} 
\right].
\label{e:s_cf_split_B}
 \ee
 \ese
 The integration over $z$ can be performed analytically, and the remaining $y$-integrals can be written as
\bse
 \be\label{e:I-integrals}
 I_1(t)=\int_0^1\diff y\; j_1(y;s,\ell)
 \csp
  I_2(t)=\int_0^1\diff y\; j_2(y;s,\ell),
 \ee
where
\be
j_1(y;s,\ell)
&=&\notag
\f{3 \left(3 s^4+2 s^2 y^2+3 y^4\right)}{32 s^5 y^2 \ell ^2 \left(y^2+\ell ^2\right)}\left[(s-y)^3 (s+y)^2-(s+y)^3(s-y) |s-y|\right]-
\\&&\notag
\f{3 \ell ^2 \left(s^2-9 y^2-5 \ell ^2\right) }{80 s^4 y \left(y^2+\ell ^2\right)}\left[(s+y)+|s-y|\right]-
\\&&\notag
\f{3 \ell ^2 \left[-32 s^4+18 y^4+40 y^2 \ell ^2+15 \ell ^4+2 s^2 \left(19 y^2+5 \ell ^2\right)\right] }
{160 s^5 y^2 \left(y^2+\ell ^2\right)}\left[(s+y)-|s-y|\right]-
\\&&\notag
\f{201 s^4+185 s^2 y^2-174 y^4 }{280 s^5 y \left(y^2+\ell ^2\right)}
\left[(s+y)^2+(s-y) |s-y|\right]+
\\&&\notag
\f{201 s^4+596 s^2 y^2-381 y^4}{560 s^4 y^2 \left(y^2+\ell ^2\right)}
\left[(s+y)^2-(s-y)|s-y|\right]+
\\&&\notag
\f{3 \left(s^2-y^2-\ell ^2\right)^2 
\left[3 s^6-s^4 \left(y^2-9 \ell ^2\right)+8s^2\ell ^4 +\left(y^2+\ell ^2\right)^2 \left(s^2-3 y^2+3 \ell ^2\right)\right] }
{32 s^5 y^2 \ell ^3 \left(y^2+\ell ^2\right)}
\times\\
&&\qquad\qquad\qquad\qquad\qquad\qquad
\;\left[\arctan\left(\f{s+y}{\ell }\right)-\arctan\left(\f{|s-y|}{\ell }\right)\right],
\\
j_2(y;s,\ell)
&=&\notag
\f{\left(y^2-9 s^2-5 \ell ^2\right) }
{20 s^2 y \left(y^2+\ell ^2\right)}
\;\left[(s+y) +|s-y|\right] +
\\&&\notag
\f{33 s^4-17 y^4+10 y^2 \ell ^2+15 \ell ^4+8 s^2 \left(y^2+5 \ell ^2\right)}
{40 s^3 y^2 \left(y^2+\ell ^2\right)}
\;\left[(s+y) -|s-y|\right] -
\\&&
\f{\left(s^2-y^2+\ell ^2\right) \left[3 s^4+3 \left(y^2+\ell ^2\right)^2+2 s^2 \left(y^2+3 \ell ^2\right)\right]}
{8 s^3 y^2 \ell  \left(y^2+\ell ^2\right)} 
\;\left[\arctan\left(\f{s+y}{\ell }\right)-\arctan\left(\f{|s-y|}{\ell }\right)\right].
\qquad
\ee
\ese
To obtain the exact correlation function, the remaining one-dimensional $y$-integrals~\eqref{e:I-integrals} can be computed numerically; for special limit cases, however, one can expand the integrands $j_{1/2}(y;s,\ell)$  and evaluate the resulting integrals  analytically. 

\paragraph*{Short-time expansion $t\to 0$.}
Expanding the integrands $j_{1/2}(y;s,\ell)$ at short times $s\ll \ell$, we find
\be
C_\mrm{S}(t)
&\simeq&\notag
\gvf\; (\gk V)^2
\left(\f{\eps}{\Gl}\right)
\frac{1}{210} 
\biggl\{
-\frac{252}{1+\ell ^2}+\frac{176 s^5}{\left(s^2+\ell ^2\right)^3}-\frac{66 s^3}{\left(s^2+\ell ^2\right)^2}-\frac{s^2 \left(489+1088 \ell ^2+423 \ell ^4\right)}{\ell ^2 \left(1+\ell ^2\right)^3}+
\\
&&\qquad\qquad\qquad\qquad\quad
s \left(\frac{489}{\ell ^2}-\frac{237}{s^2+\ell ^2}\right)+
\frac{3 \left(163 s^2-84 \ell ^2\right) 
\left[\text{arctan}(\ell )-\text{arctan}\left({\ell }/{s}\right)\right]}{\ell ^3}
\biggr\},
\ee
where $\ell=\gl/\Gl$ and $s:=t V/\Gl$.

\paragraph*{Large-time (small $\ell$) approximation and thermodynamic limit.}
To obtain an analytically tractable approximation of the autocorrelation function that can be used to determine the thermodynamic limit, we note that,  at large times, we can approximate $\ell \simeq 0$ in the denominators of the integrals 
Eqs.~\eqref{e:s_cf_split_A} and~\eqref{e:s_cf_split_A}. With this simplification, the $z$-integral can be computed more easily, yielding  for $0<s<1$
\be
C_\mrm{S}(t)\simeq
\f{3}{2}\;\gvf\;(\gk V)^2 \left(\f{\eps}{\Gl}\right)
\int_0^1\diff y
\;\left[-\frac{16 \left(7 s^2 y-6 y^3\right)}{35 s^5}\right]\Gt(s-y)
+
\left[\frac{8 \left(-9 s^2+7 y^2\right)}{35 y^4}\right]\Gt(y-s),
\label{e:dip_C_approx}
\ee
where $\Gt(x)$ is the unit step function, defined by $\Gt(x):=0, x<0$ and $\Gt(x):=1, x\ge 0$.  In principle, the remaining $y$ integral can readily be evaluated to obtain the long-time behavior of $C_\mrm{S}(t)$. However, the resulting expression diverges at short times,   since we let the cut-off $\ell\to 0$. To avoid this divergence and mimic the effect of the short-distance cut-off $\ell$, we may replace the  lower integral boundary in Eq.~\eqref{e:dip_C_approx} by a regularization parameter $\ell_*$~\footnote{For numerical simulations, it is usually more convenient to regularize divergent flow fields. For analytical calculations, it is often advisable to consider the corresponding non-regularized flow fields and to regularize divergences by adapting the integral boundaries.}, i.e., we compute
\be
C_\mrm{S}(t)\simeq
\f{3}{2}\;\gvf\;(\gk V)^2 \left(\f{\eps}{\Gl}\right)
\int_{\ell_*}^1\diff y
\;\left\{
\left[-\frac{16 \left(7 s^2 y-6 y^3\right)}{35 s^5}\right]\Gt(s-y)
+
\left[\frac{8 \left(-9 s^2+7 y^2\right)}{35 y^4}\right]\Gt(y-s)
\right\},
\label{e:dip_C_approx_2}
\ee
which gives
\bse\label{e:dip_C_approx_3}
\be\label{e:dip_C_approx_3a}
C_\mrm{S}(t)\simeq
\gvf\;(\gk V)^2 \left(\f{\eps}{\Gl}\right)
\f{12}{5}
\begin{cases}
-1+\frac{1}{\ell_*}+\frac{3 s^2}{7}-\frac{3 s^2}{7 \ell_*^3}, &\ell_*\ge s,\\
-1-\frac{3 \ell_*^4}{7 s^5}+\frac{\ell_*^2}{s^3}+\frac{3 s^2}{7}, &\ell_*<s<1. 
\end{cases}
\ee
The regularization parameter $\ell_*$ can be determined from the condition $C_\mrm{S}(0)\overset{!}{=}\lan |\bs v|^2\ran$, by using the exact result~\eqref{e:dipol_moments_2} for the second moment, yielding  
\be
\ell_*\simeq \frac{4}{\pi }  \ell.
\ee 
\ese
The result~\eqref{e:dip_C_approx_3} becomes exact at sufficiently large times $s\to 1$;  it does, however, also provide a useful approximate description at intermediate and small times. In particular, the second line in  Eq.~\eqref{e:dip_C_approx_3a} implies that, for a finite system, the autocorrelation function becomes negative after a certain time $t_0$, which can be estimated as
\be\label{e:C_zero}
t_0\overset{\ell\ll 1}{\simeq}\left(\f{4}{\pi} \f{\gl}{\Gl} \right)^{2/3}\left( \f{\Gl}{V}\right).
\ee
Physically, this due to the dipolar flow field structure: If a dipolar swimmer passes a fixed point in the fluid, the flow at this point will reverse its sign (direction) after certain time.

However, as evident from Eq.~\eqref{e:C_zero}, the  negative correlation region vanishes for $\Gl\to \infty$, since the zero $t_0$ of $C_\mrm{S}(t)$ moves to $\infty$ in this limit. More precisely, by taking  the thermodynamic limit of Eq.~\eqref{e:dip_C_approx_3}, we find that
\be\label{e:dip_C_approx_TDL}
C_\mrm{S}(t)\simeq
\gvf\;(\gk V)^2 \left(\f{\eps}{\gl}\right)
\f{3\pi}{5} 
\begin{cases}
1- \frac{3 t^2}{7 \tau_\gl^2} & t  \leq \tau_\gl:=\f{4}{\pi}\f{\gl}{V} , \\
 \frac{ \tau_\gl^3}{t^3} - \frac{3 \tau_\gl^5}{7t^5} & t> \tau_\gl.
\end{cases}
\ee
Thus, the velocity field  autocorrelation function in a dipolar swimmer suspension decays asymptotically as $t^{-3}$, which  is different from
the $t^{-1}$-decay found earlier for the power-law model with $n=2$, see Eq.~\eqref{e:alpha=2_C_longtime}.

\subsection{Spatial mean square displacement of a tracer particle}

The approximate result for the velocity autocorrelation can be used to obtain an upper bound for the swimmer contribution to the mean square displacement (MSD) of the tracer particles.  Considering an initial tracer position $\bs x(0)=\bs 0$ and noting that $\lan \bs u_N(\tau',\bs 0)\bs u_N(\tau,\bs 0)\ran=C(|\tau'-\tau|)$, we find
\be
\lan [\bs x(t)]^2\ran
&=&\int_0^t\diff \tau' \int_0^t\diff\tau \;\lan \bs u_N(\tau',\bs x(\tau'))\bs u_N(\tau,\bs x(\tau))\ran
\notag\\
&\lesssim&
\int_0^t\diff \tau' \int_0^t\diff\tau \;\lan \bs u_N(\tau',\bs 0)\bs u_N(\tau,\bs 0)\ran
\notag\\
&=&\label{e:msd_approx}
\int_0^t\diff \tau' \int_0^t\diff\tau \;C(|\tau'-\tau|).
\ee
The second line reflects, roughly speaking,  the assumption that correlations are spatially homogeneous and decaying with distance, i.e.,
\be
\lan \bs u_N(\tau',\bs x')\bs u_N(\tau,\bs x)\ran 
\le 
\lan \bs u_N(\tau',\bs x)\bs u_N(\tau,\bs x)\ran
=
\lan \bs u_N(\tau',\bs 0)\bs u_N(\tau,\bs 0)\ran
\ee
\par
Changing integration variables,  $\tau\mapsto\gt:=\tau'-\tau$, and, subsequently, $\tau'\mapsto\nu:=\tau'-t$, we may rewrite Eq.~\eqref{e:msd_approx} as
\be
\lan [\bs x(t)]^2\ran
&\simeq&
\int_{-t}^0 \diff \nu \int_{\nu}^{\nu+t} \diff\gt \;C(|\gt|).
\ee
Inserting the  approximate result~\eqref{e:dip_C_approx_TDL} for the fluid autocorrelation function in the thermodynamic limit, we find
\begin{eqnarray}
\lan [\bs x(t)]^2\ran
&\simeq&
6\,\gvf\; \gk^2 V\eps\; t
\begin{cases}
\frac{2 t  }{5\tau_\gl}-\frac{t^3 }{35 \tau_\gl^3}, &  t<\tau_\gl:=\f{4}{\pi}\f{\gl}{V},\\
1- \f{\tau_\gl}{t}+\frac{2  \tau_\gl^2}{5 t^2 }-\frac{\tau_\gl^4}{35 t^4},   &  t\ge\tau_\gl.
\end{cases}
\end{eqnarray}
Thus, tracer diffusion in a dilute suspension of dipolar swimmer  is ballistic at short times  $t\ll \tau_\gl$ and normal at large times $t\gg \tau_\gl$.

\section{Numerical methods}
\label{s:numerics}
In our computer simulations, we directly integrate the  Langevin equations
\bse\label{e:sim_model}
\be
\label{e:sim_model_tracer}
\dot{\bs x}(t) &=& \bs{u}_N(t, \bs{x}(t)) + \sqrt{2D_0}\;\boldsymbol{\xi}(t),\\
\bs u_N(t, \bs x)&=&\sum_{\gs=1}^N\bs u(\bs x|\Gc^\gs(t))
\csp
\Gc^\gs(t)=(\bs X^\gs(t),\bs V^\gs (t)),\\
\bs X^\gs(t)&=&\bs X_0^\gs+t {\bs V}^\gs_0,
\csp \gs=1,\ldots, N,
\label{e:sim_model_swimmer}
\ee
\ese
where $\bs{\xi}(t)$ is Gaussian white noise, to obtain the velocity distribution at a given point in the fluid  and velocity autocorrelaton functions (for $D_0=0$), and 
the tracer position PDF ($D_0>0$). The initial positions and velocities $\Gc^\gs(0)=(\bs X^\gs_0,\bs V^\gs_0)$ of the swimmers were sampled from the distribution~\eqref{e:S_PDF}.

\paragraph*{Particle deletion \& insertion.--}
Using the Euler method, we simulate an ``ideal gas'' of active particles (swimmers), which move according to Eq.~\eqref{e:sim_model_swimmer} through sphere of radius $\Gl$. This sphere is always relative to the passive tracer, whose position evolves according
to Eq.~\eqref{e:sim_model_tracer}. If an active particle leaves the sphere, we immediately delete it. To restore detailed
balance, we continually insert new active particles at the boundary of the sphere. The number of insertions
per time step is drawn from a Poisson distribution $f(j) = \nu^{j} e^{-\nu} / j!$, where $\nu$ is the mean
number of insertions during $\Delta t$. We obtain equilibrium by setting $\nu$ equal to the mean number of
deletions during $\Delta t$, which may be estimated from the kinetic theory of gases \cite{2007Ca} as
$\nu = \frac{3 N V}{4 \Gl} \Delta t$. For each insertion, it is necessary to bias the orientation of an
active particle so that its probability distribution satisfies $p(\hat{\bs \Go}^\gs | \hat{\bs R}^\gs)
\propto \hat{\bs \Go}^\gs \hat{\bs R}^\gs$ normalized over the solid angle of a hemisphere with inward surface
normal $\hat{\bs R}^\gs$. We achieve that by uniformly choosing a position ${\bs R}^\gs$ at a distance $\Gl$
relative to the tracer, then choosing an orientation from $p(\hat{\bs \Go}^\gs | \hat{\bs R}^\gs)$. Generally, our
procedure achieves numerical accuracy if $\Gl$ is large, and ensures that a suspension of ballistic particles remains
homogeneous and isotropic with mean population $N$.  A comparison with the exact results for time-dependent velocity-correlations verifies the validity of this approach.

\paragraph*{GPU implementation.--}
Resolving the tail of a probability distribution can be a computationally expensive task, even
for stochastic processes that are relatively simple. In our numerical calculations, further
difficulties arise from having to create and maintain an active suspension unique to each tracer.
This process would not be possible in a reasonable amount of time on a traditional computer. We therefore
implemented parallelized simulations on a graphics processing unit (GPU) using NVIDIA's Compute Unified Device
Architecture (CUDA). Compared to a single-core CPU, GPU code yields substantial speed-ups (up to a factor of a
few hundreds). However, our longest simulation of 4194304 trajectories still took 14 days on a GPU.

\paragraph*{Acknowledgements}
This work was supported by the Natural Sciences and Engineering Research Council of Canada (I.~M.~Z.),
and the ONR, USA (I.~M.~Z. and J.~D.).

